\documentclass[11pt, letterpaper]{article}

\usepackage[bookmarksnumbered,bookmarksopen,colorlinks,citecolor=blue,linkcolor=blue]{hyperref}

\usepackage{multirow} 
\usepackage{subfigure} 
\usepackage{natbib} 

\usepackage{graphicx}

\usepackage{JASA_manu}

\usepackage{ifthen}

\usepackage{amsfonts}
\usepackage{mathrsfs}
\usepackage[centertags]{amsmath}
\usepackage{amssymb}
\usepackage{amsthm}

\usepackage{color}

\usepackage{longtable}

\usepackage{verbatim}
\usepackage{listings}

\usepackage{xr}
\definecolor{impSen}{rgb}{1.00,0.00,0.00}
\definecolor{impWord}{rgb}{1.00,0.00,0.00}
\definecolor{impFor}{rgb}{1.00,0.00,0.00}
\definecolor{impPar}{rgb}{1.00,0.00,0.00}

\definecolor{Questions}{rgb}{0.00,0.50,0.00}
\definecolor{Tasks}{rgb}{0.00,0.50,0.00}
\definecolor{Notes}{rgb}{0.00,0.50,0.00}

\definecolor{light}{gray}{.8} 

 \theoremstyle{plain}
 \newtheorem{thm}{Theorem}
 \newtheorem{cor}[thm]{Corollary}
 \newtheorem{lem}[thm]{Lemma}
 \newtheorem{prop}[thm]{Proposition}

 \theoremstyle{definition}
 \newtheorem{defn}{Definition}
 \newtheorem{conj}{Conjecture}[section]
 \newtheorem{exmp}{Example}[section]

 \theoremstyle{remark}
 \newtheorem{rem}{Remark}[section]

 \numberwithin{equation}{section}

 \ifthenelse{1=2}
     {
     \newenvironment{TangThm}{\begin{framed}\begin{thm}}{\end{thm}\end{framed}}
     \newenvironment{TangCor}{\begin{framed}\begin{cor}}{\end{cor}\end{framed}}
     \newenvironment{TangLem}{\begin{framed}\begin{lem}}{\end{lem}\end{framed}}
     \newenvironment{TangProp}{\begin{framed}\begin{prop}}{\end{prop}\end{framed}}

     \newenvironment{TangDefn}{\begin{framed}\begin{defn}}{\end{defn}\end{framed}}
     \newenvironment{TangConj}{\begin{framed}\begin{conj}}{\end{conj}\end{framed}}
     \newenvironment{TangExmp}{\begin{framed}\begin{exmp}}{\end{exmp}\end{framed}}

     \newenvironment{TangRem}{\begin{framed}\begin{rem}}{\end{rem}\end{framed}}
     }
     {
     \newenvironment{TangThm}{\begin{thm}}{\end{thm}}
     
     \newenvironment{TangLem}{\begin{lem}}{\end{lem}}
     \newenvironment{TangProp}{\begin{prop}}{\end{prop}}

     \newenvironment{TangRem}{\begin{rem}}{\end{rem}}
     }

\newcommand{\kls}{\left(} 
\newcommand{\krs}{\right)}
\newcommand{\klm}{\left[} 
\newcommand{\krm}{\right]}
\newcommand{\kll}{\left\{} 
\newcommand{\krl}{\right\}}

\newcommand{\WC}{\overset{d}{\rightarrow}}
\newcommand{\argmin}{\text{Argmin}}

\usepackage{fancybox} 

{\begin{center}\begin{Sbox}\begin{minipage}{5in}
\emph{\textcolor{Questions}{\textbf{QUESTIONS:}}}}%
{\end{minipage}\end{Sbox}\fbox{\TheSbox}\end{center}}

{\begin{center}\begin{Sbox}\begin{minipage}{5in}
\emph{\textcolor{Tasks}{\textbf{TASKS:}}}}%
{\end{minipage}\end{Sbox}\fbox{\TheSbox}\end{center}}

{\begin{center}\begin{Sbox}\begin{minipage}{5in}
\emph{\textcolor{Notes}{\textbf{NOTES:}}}}%
{\end{minipage}\end{Sbox}\fbox{\TheSbox}\end{center}}

\newcommand{\DD}{\ensuremath{\mathbb{D}}} 
\newcommand{\CherDist}{\ensuremath{\mathcal{Z}}} 
\newcommand{\ZZ}{\ensuremath{\mathcal{Z}}}

\usepackage{framed}
\definecolor{shadecolor}{gray}{.75}

\usepackage{lipsum}

\usepackage[margin=1in]{geometry}

\usepackage{authblk}
\date{}
\begin{document}
\title {Two-Stage Plans for Estimating a Threshold Value  \\ of a Regression Function}
\author[1]{Runlong Tang}
\author[2]{Moulinath Banerjee}
\author[2]{George Michailidis}
\author[2]{Shawn Mankad}
\affil[1]{Department of Operations Research and Financial Engineering, Princeton University}
\affil[2]{Department of Statistics, University of Michigan}
\renewcommand\Authfont{\scshape\small}
\renewcommand\Affilfont{\itshape\small}
\setlength{\affilsep}{0em}

\maketitle


\begin{center}
\textbf{Abstract}
\end{center}

This study investigates two-stage plans based on nonparametric procedures for estimating an inverse regression function at a given point.
Specifically, isotonic regression is used at stage one to obtain an initial estimate followed by another round of isotonic regression in the vicinity of this
estimate at stage two. It is shown that such two stage plans accelerate the convergence rate of one-stage procedures and are superior to existing two-stage procedures that use local parametric approximations at stage two when the available budget is moderate and/or the regression function is `ill-behaved'. Both Wald and Likelihood Ratio type confidence intervals for the threshold value of interest are investigated and the latter are recommended in applications due to their simplicity and robustness.
The developed plans are illustrated through a comprehensive simulation study and an application to car fuel efficiency data.



%


\pagenumbering{arabic}  



\section{Introduction}\label{sec1}

Threshold estimation is a canonical statistical estimation problem with numerous applications in 
science and engineering. Here is an interesting motivating example. In recent years,
an important consideration for both car manufacturers and potential buyers in 
the United States is the fuel efficiency of the vehicle, expressed in miles per gallon (MPG).
The National Highway Traffic Safety Administration (NHTSA) regulates
the Corporate Average Fuel Economy (CAFE) standards to encourage automobile manufacturers
to improve the average fuel efficiency of their fleets of vehicles.
The CAFE standard for 2012 models is 29.8 MPG, set to increase to 34.3 MPG in 2016 according to 
the Environmental Protection Agency rule that came into effect in August 2012.
To encourage higher fuel efficiency, manufacturers are subject to a penalty if the average FE of their
fleets falls below the CAFE standard. Moreover, a so-called gas guzzler tax is imposed on cars with
low FE in accordance with the US Energy Tax Act of 1978. 

\begin{figure}[!h]
  \begin{center}
  \includegraphics[scale=0.43]{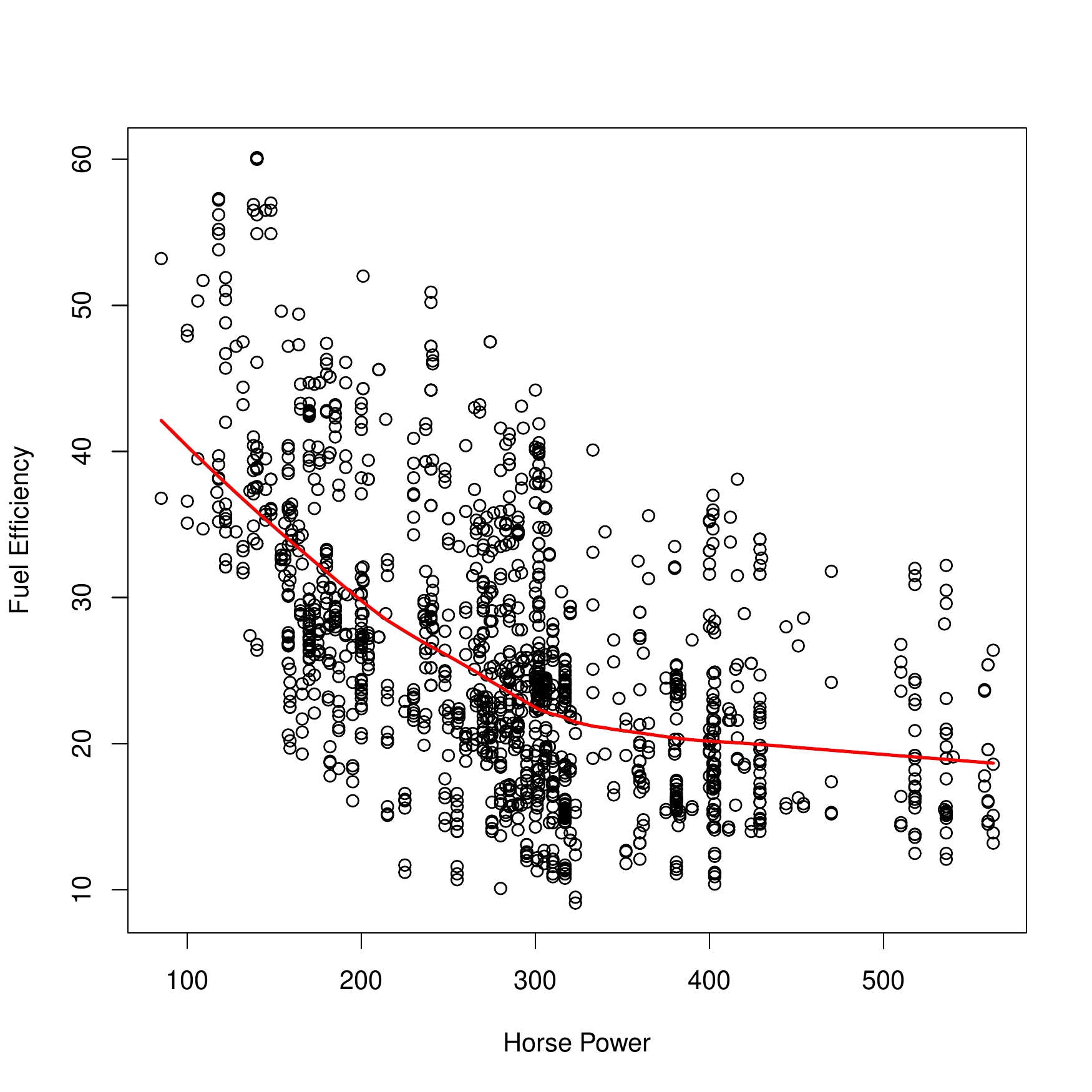}
   \includegraphics[scale=0.43]{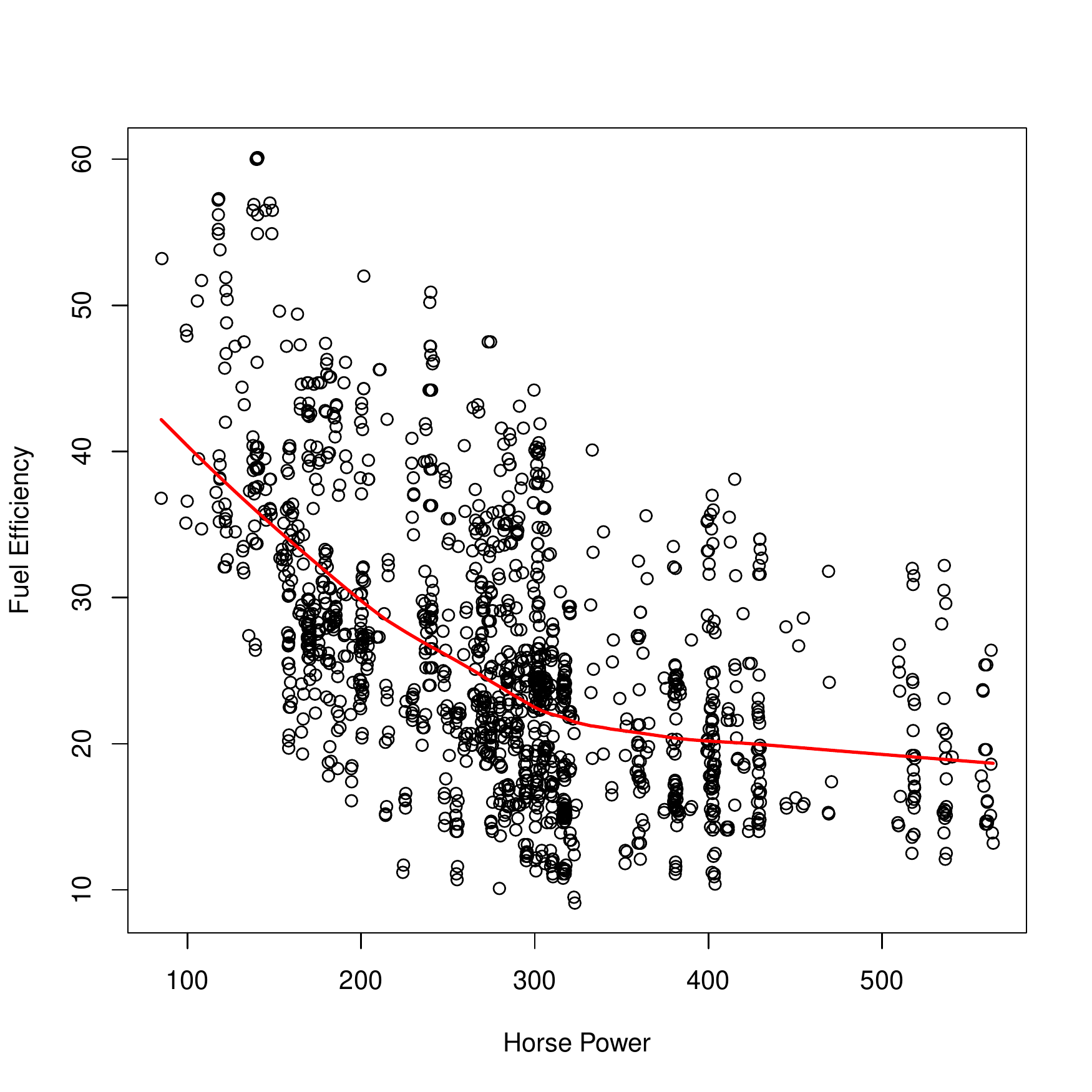}
  \caption
    {Scatterplots of the relationship between horse power and fuel efficiency of naturally aspirated vehicles; $FE=m(HP)$. 
The right panel shows the data with jittered horsepower to create a unique horsepower for every observation.
   }
  \label{paper:FG:DataAnalysis:HorsePowerCombinedFuelEfficiencyAirAspiration}
  \end{center}
\end{figure}

The data on fuel efficiency as a function of the vehicle's horse power, which is a key component, for the
2012 models is shown in Figure \ref{paper:FG:DataAnalysis:HorsePowerCombinedFuelEfficiencyAirAspiration} (for a detailed discussion of the data and CAFE standards,
see section 4). An expected decreasing relationship is observed and
it is of interest to identify the horse power threshold at which the fuel efficiency meets the current
CAFE, as well as the 2016 standard.

The data plot indicates that fitting a precise parametric model may be challenging, while it is rather
straightforward to fit a monotonically decreasing nonparametric one and obtain the fuel efficiency
threshold for the target values of $\sim 30$ and $\sim 34$. However, it is also desirable to assign
a confidence interval around the estimate and an interesting question addressed in this paper is
whether an {\em adaptive procedure} can lead to improved precision for such threshold estimates.

The topic of using adaptive procedures in a {\em design setting}
for threshold estimation models has been recently studied in the literature \citep{lan2009,tang2011}.
The basic model considered is $Y = m(X) + \epsilon$,
where the design point $X$ takes values in $[a,b]$,
the regression function $m$ is monotone, for the sake of presentation henceforth assumed non-decreasing,
and the random error $\epsilon$ has mean $0$ and finite variance $\sigma^{2}$.
The quantity to be estimated is a threshold $d_0$, which in \citep{lan2009} corresponded to a change-point 
(i.e. $m(X)=\alpha_0 1(x\leq d_0) +\beta_0 1(x>d_0)$ with unknown constants $\alpha_{0}$ and $\beta_{0}$),
whereas in \cite{tang2011} to $d_0=m^{-1}(\theta_0)$ for some prespecified $\theta_0$.

The employed two-stage adaptive procedure in \cite{lan2009} and \cite{tang2011} works as follows: (i) in the first stage,
it utilizes a portion $p$ of the design budget to obtain an initial estimate of $d_0$, (ii)  in the second stage, 
the remaining portion ($1-p$) of the budget is used
to obtain more sample points in a small neighborhood of that estimate; (iii) finally, an improved estimate
based on the second-stage data is constructed. 
This more intense ``zoom-in" sampling leads to {\em accelerated}
convergence rates of the second stage estimators for $d_0$ compared to the standard ones that use all the data
in one shot. Specifically, for the change point problem, the rate can be accelerated from $n$
to almost $n^2$ (up to a logarithmic factor) as in \citep{lan2009}, while for the inverse regression problem, from $n^{1/3}$
to $n^{1/2}$ by employing a local linear approximation \citep{tang2011}, where $n$ denotes the total budget
available. Hence, tighter confidence intervals can be constructed that have the correct nominal coverage
with the same budget as standard one-stage procedures, or alternatively one can reduce the design budget and
still have good quality confidence intervals.

In this paper, given our motivating data application, 
we focus on the second problem, namely that of estimating the inverse regression function
at a prespecified point $\theta_0$. This is closely related to dose response \textcolor{blue}{\citep{Rosenberger2002}} 
and statistical calibration studies
\textcolor{blue}{\citep{Osborne1991}} (for additional references see \cite{tang2011}).
As mentioned above, a strategy that obtains a first stage estimate
using isotonic regression, followed by a local linear approximation, gives a consistent second stage
estimator that achieves the parametric rate $\sqrt{n}$. However, the success of this strategy \emph{heavily hinges
upon the (approximate) linearity of the regression function $m(\cdot)$ in the vicinity of $d_0$}. Small departures from
linearity do not adversely affect the results (especially when the budget is large enough to allow for significant
``zooming-in" at the second stage), but severe departures are a totally different matter as illustrated next.

\begin{figure}[!h]
\begin{center}
  \includegraphics[scale=0.46]{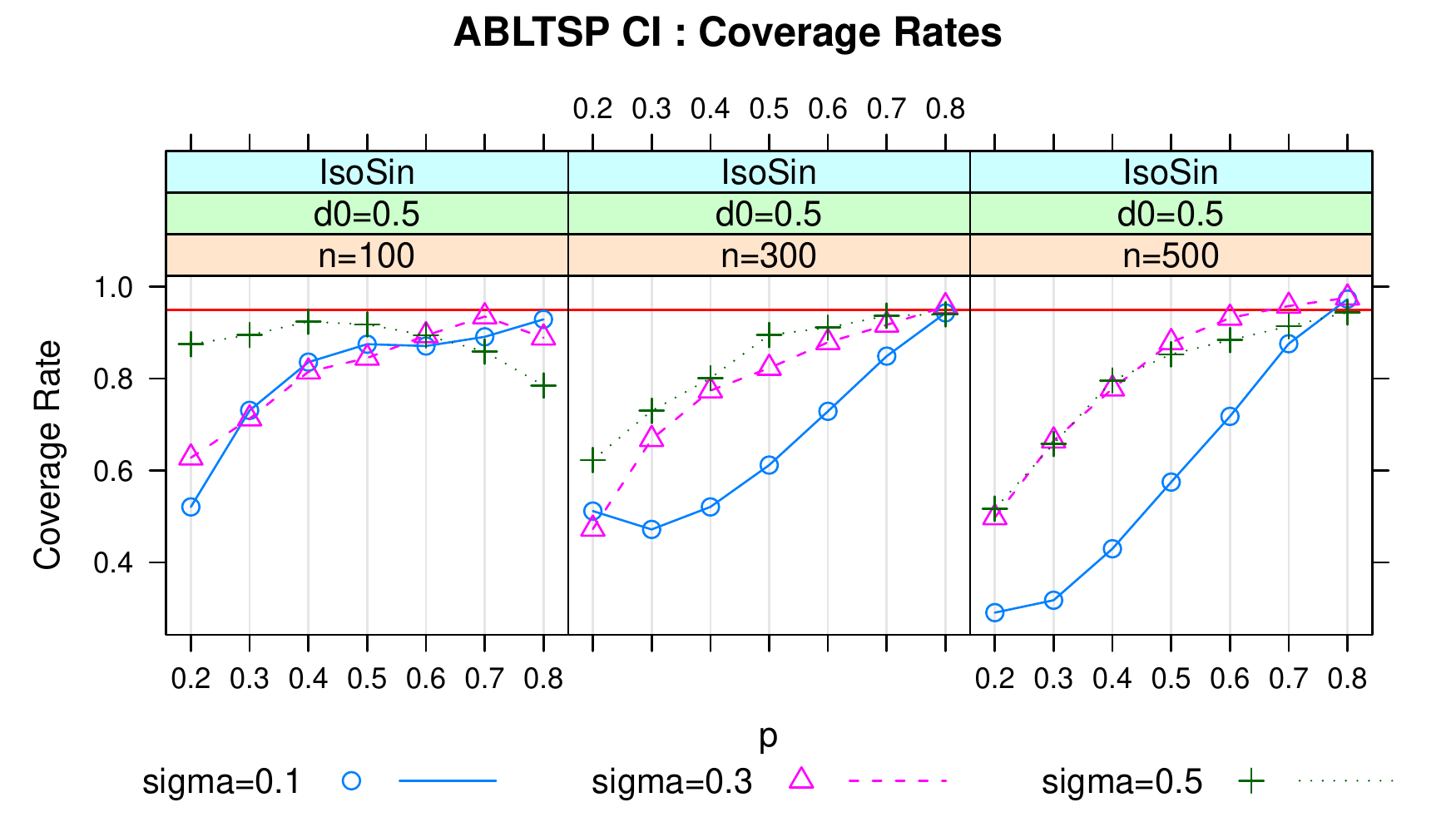}
  \includegraphics[scale=0.46]{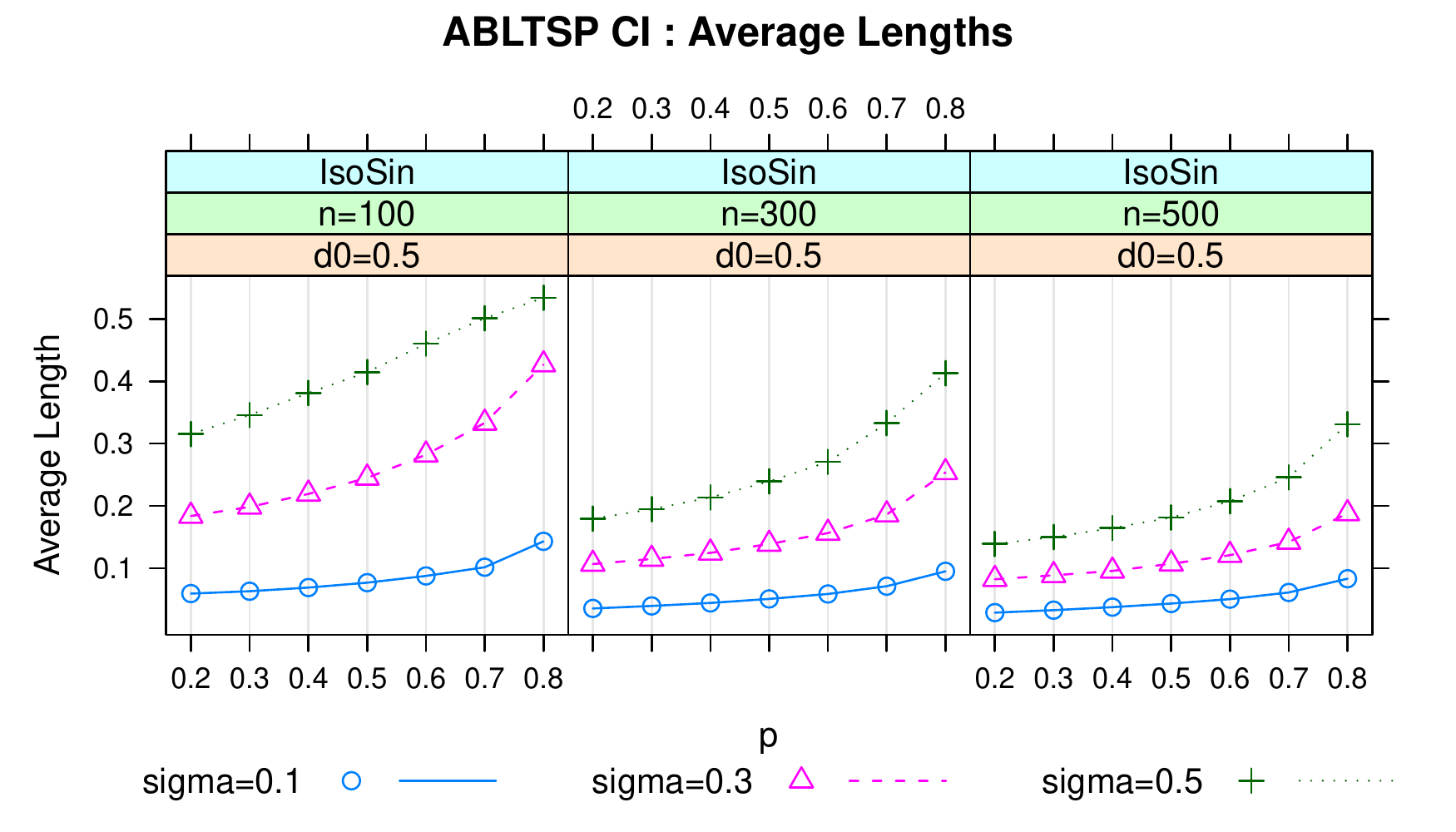}
  \caption{\label{illustration1} The left panel shows the coverage rates
of the 95\% confidence intervals using a local linear approximation for $d_{0} = 0.5$ with different sample sizes, and noise levels. 
The right panel shows the corresponding average lengths of the intervals.
}
\end{center}
\end{figure}

Consider a monotone regression function exhibiting strong nonlinearity at $d_0=0.5$, for example, given by
$m(x) = (1/40) \sin(6\pi x) + 1/4 + (1/2)x + (1/4)x^2$ with $x\in [0,1]$
(see the left panel of Figure \ref{paper:fg:regressionFunctions:derivative}
for its plot). The coverage rates and average lengths of the confidence intervals obtained from
the two-stage adaptive strategy based on isotonic regression and a local linear approximation for
selected total budget sizes ($n=100, 300, 500$), varying portions $p$ allocated to the first stage
and different noise levels ($\sigma=0.1, 0.3, 0.5$) are depicted in Figure \ref{illustration1}. 

It can be seen that
for the majority of portions $p$, the confidence intervals constructed from this adaptive strategy fail miserably in terms of
coverage rates and exhibit relatively large lengths.  
Of interest is the fact that for large $p$, 
the coverage rates indeed approach the nominal level.
In practice, however, it is not possible to choose an appropriate $p$ without prior information on $m$.
Further, even for large $p$'s, the confidence intervals are excessively wide, especially for large noise levels.

\begin{figure}[!h]
\begin{center}
  \includegraphics[scale=0.45]{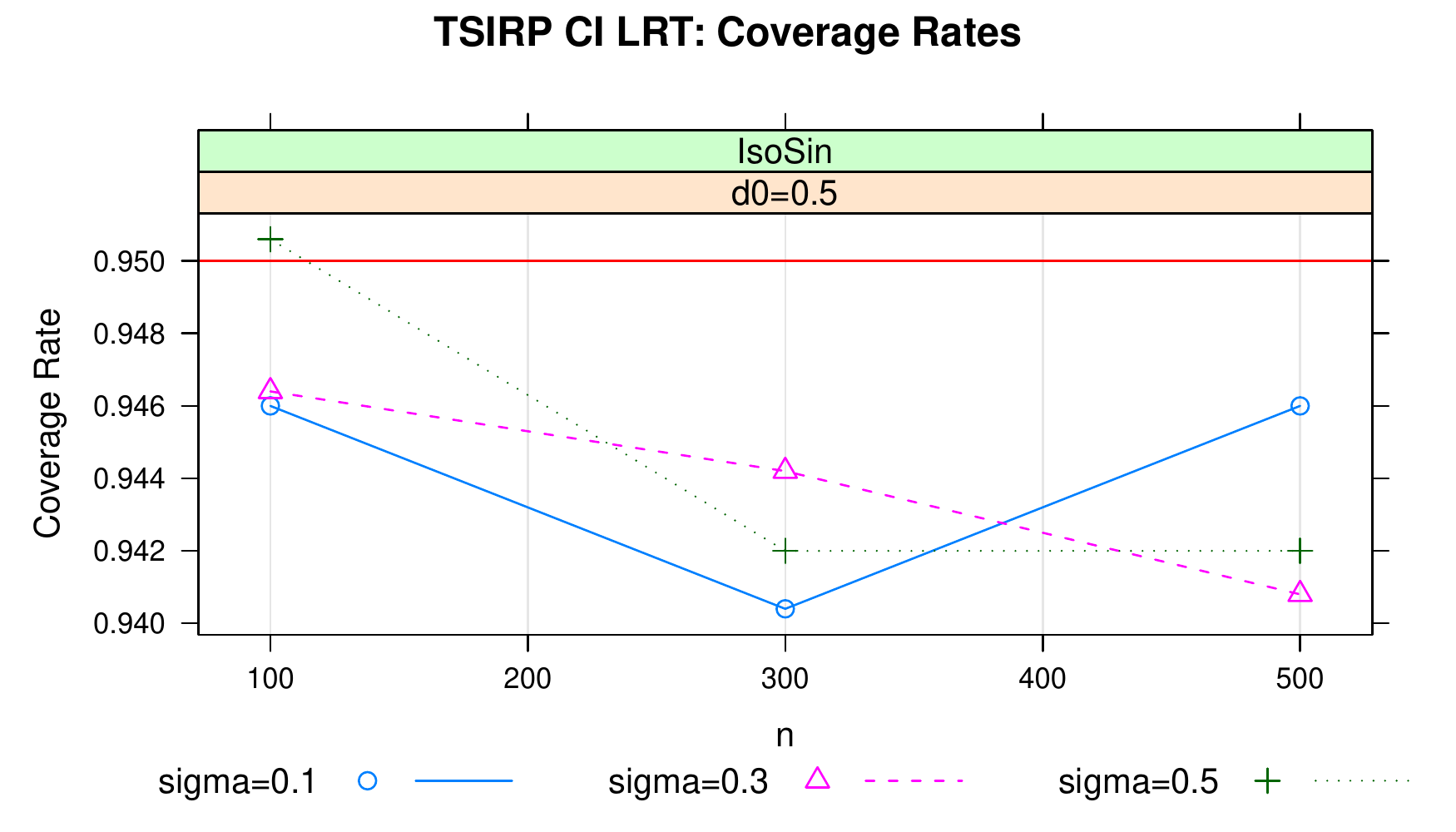} 
  \includegraphics[scale=0.45]{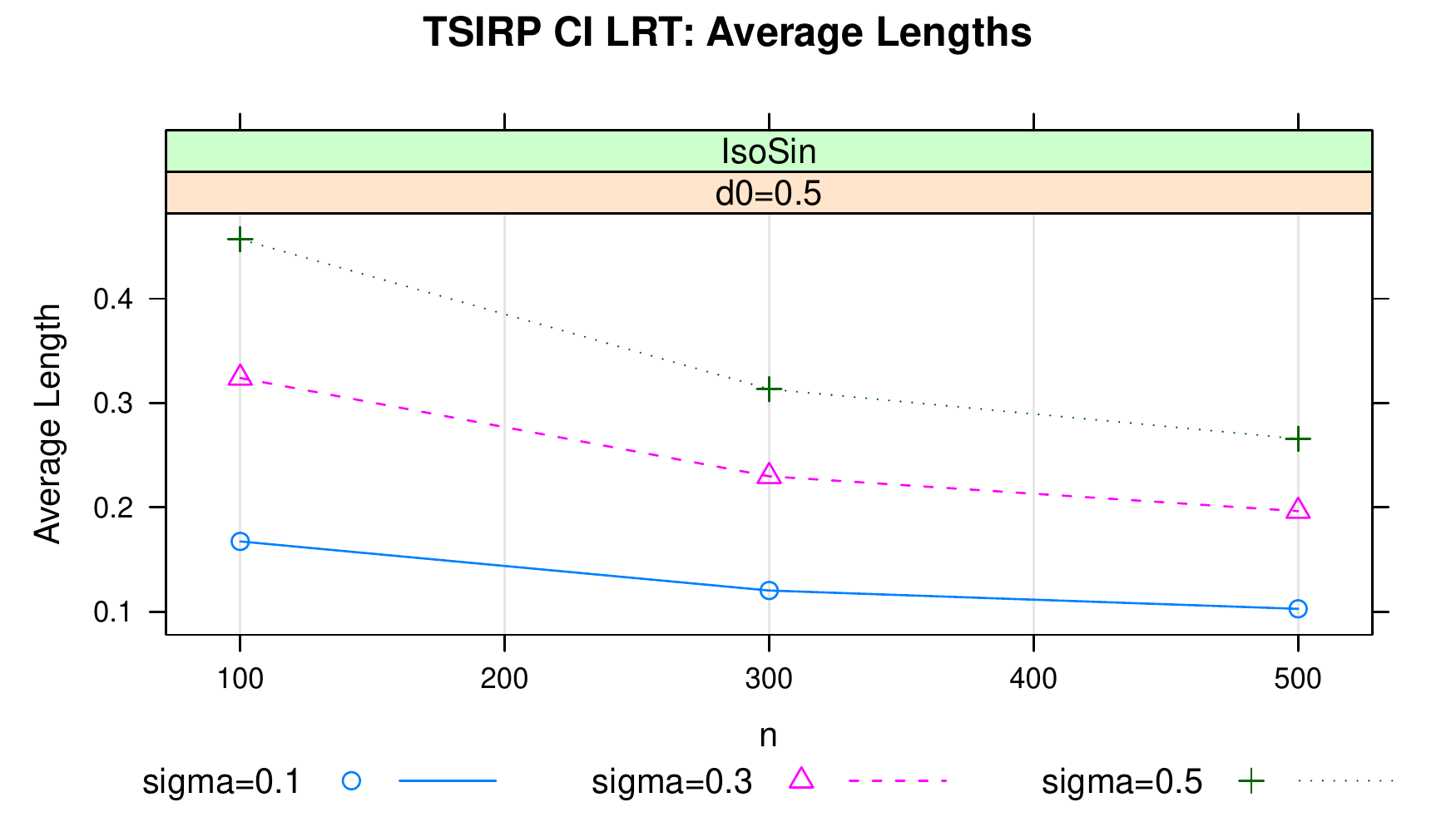}
  \caption{\label{illustration2} The left panel shows the coverage rates
of the 95\% confidence intervals using a two stage procedure for $d_{0} = 0.5$ with different sample sizes, and noise levels. The right panel shows the corresponding average lengths of the intervals.}
\end{center}
\end{figure}

In contrast, a two stage adaptive strategy based on employing isotonic regression at both stages,
which will be fully developed in this paper, overcomes
these difficulties. Such a strategy would be also desirable for the motivating data application,
due to high variability in the vicinity of the CAFE thresholds, as seen in the scatterplots of 
Figure \ref{paper:FG:DataAnalysis:HorsePowerCombinedFuelEfficiencyAirAspiration}.
In Figure \ref{illustration2}, the coverage rates and average lengths using our new strategy 
are shown for the same settings as above, but with $p=1/4$ (for more on this universal choice of $p$ see Section \ref{sec2}).
It can be seen that this wholly nonparametric strategy overcomes the previous difficulties, proves robust to
the level of local nonlinearity of the regression function $m$ and, as argued in Section \ref{sec2}, is easy to
implement.

The remainder of the paper is organized as follows: in Section \ref{sec2} the adaptive procedure is
introduced and the main results presented. Section \ref{sec3} presents extensive simulations results, while
an interesting application of the methodology to fuel efficiency data is shown in Section \ref{TSPInvFun:paper:DataAnalysis}. 
Section 5 concludes. Proofs are sketched in the appendix. 

\section{Two Stage Adaptive Procedures}
\label{sec2}

\subsection{An Overview of the Isotonic Regression Procedure}\label{sec:summary}

We provide a brief description of the one-stage isotonic regression procedure (OSIRP).
Specifically, given $n$ fixed or random design points $\{X_{i}\}_{i=1}^{n}$ in $[a,b]$ distributed
according to a continuous design density $g$ 
and the corresponding responses $\{Y_{i}\}_{i=1}^{n}$,
obtained from the proposed model,  the isotonic regression estimate of $m(\cdot)$ is given by
\begin{equation} \label{OSIRP:IsotonicRegression}
m_{I}(x)=
m_{1}^{\star} 1\{x\in[a,X_{1}]\} + \sum_{i=1}^{n-1} m_{i}^{\star} 1\{x\in[X_{i},X_{i+1})\} + m_{n}^{\star} 1\{x\in[X_{n},b]\}
\end{equation}
where 
    $\{m_i^{\star}\}_{i=1}^n = 
    \underset{m_1 \leq m_2 \leq \ldots \leq m_n}{\argmin}\sum_{i=1}^n\,(Y_i - m_i)^2$.
This minimizer exists uniquely, has a nice geometric characterization as the slope of the greatest convex
minorant of a stochastic process and is readily computable using the
pool adjacent violators algorithm (PAVA)
(see, for
example, Robertson et. al. (1988)).
Then, for a prespecified value $\theta_{0}\in (m(a), m(b))$,
the one-stage isotonic regression estimator of $d_{0}$
is defined by
\begin{equation}\label{OSIRP:dI}
    d_{I}=m_{I}^{-1}(\theta_{0})=\inf \{x\in[a,b]: m_{I}(x)\geq \theta_{0}\},
\end{equation}
where $\inf\{\emptyset\}=b$.
Under mild conditions on the regression function and the design density, namely 
\newline
{\bf Assumption A:}  $m$ is once continuously differentiable in a neighborhood of $d_{0}$ 
with positive derivative $m'(d_{0})$ and $g$ is positive and continuous at $d_{0}$, 
\newline
the asymptotic distribution of $d_I$ is given by (see \cite{tang2011}):
 \begin{equation}\label{OSIRP:Thm:d0:Wald:Equ}
  n^{1/3}(d_{I} - d_{0}) \overset{d}{\rightarrow}
  C_{d_{I}}g(d_{0})^{-1/3} \mathcal{Z},
 \end{equation}
 where $C_{d_{I}}=\left(4\sigma^{2}/m'(d_{0})^{2}\right)^{1/3}$
  and $\mathcal{Z}$ follows the standard Chernoff distribution (\citet{Groeneboom2001}).
This result can be used to construct a $1-\alpha$ Wald-type confidence interval for $d_0$:
\begin{equation*}
    \left[d_I \pm n^{-1/3}\,\widehat{C_{d_I}}\,\widehat{g(d_0)}^{-1/3}\,q(\mathcal{Z}, 1-\alpha/2)\right],
\end{equation*}
where the hats denote consistent estimates and $q(\xi,\tau)$ is the lower $\tau$'th quantile of a random variable $\xi$. 

An alternative is to construct confidence intervals through likelihood ratio (LR) testing.
Specifically, the hypotheses of interest are
\begin{equation}\label{OSIRP:LRT:Hypotheses}
    H_{0}: m^{-1}(\theta_{0}) = d_{0}
    \leftrightarrow
    H_{a}: m^{-1}(\theta_{0}) \not = d_{0}.
\end{equation}
Then, the LR test statistic is given by
\begin{equation}\label{OSIRP:LRT}
  2\log\lambda_{I}
     =2\log\lambda_{I}(d_{0})
     =2 \klm l_{n}(m_{I}, \hat{\sigma})-l_{n}(m_{Ic}, \hat{\sigma}) \krm,
\end{equation}
where
$  l_{n}(m, \sigma)=-(2\,\sigma^{2})^{-1}\sum_{i=1}^{n}(Y_{i}-m(X_{i}))^{2}$, $m_{Ic}$ is
the constrained isotonic regression of $m$ under $H_{0}$ and $\hat{\sigma}$ a consistent estimate of $\sigma$.
It is known that $m_{Ic}$ uniquely exists (see \cite{Banerjee2000c}).
The asymptotic distribution of $2\log\lambda_{I}$ under $H_{0}$ is given in \citep{Banerjee2009}:
$2\log\lambda_{I} \overset{d}{\rightarrow} \mathbb{D}$,
where $\mathbb{D}$ is a `universal' random variable not depending on the parameters of the model (\citet{Banerjee2001}).
This result allows us to construct a $1-\alpha$ LR-type confidence region for $d_{0}$:
\begin{equation}\label{OSIRP:Thm:d0:LRT:CI}
 \{x\in[a,b]: 2\log\lambda_{I}(x)\leq q(\mathbb{D}, 1-\alpha)\}.
\end{equation}
The  LR-type confidence region can be shown to be an interval and is typically \emph{asymmetric} around $d_{I}$, unlike the
Wald-type one. Its main advantage is that only $\sigma$ needs to be estimated for
its construction, whereas for the Wald confidence interval, estimation of $m'(d_0)$ is
also needed, a significantly more involved task.
\begin{TangRem} 
The use of the term LR statistic in connection with \ref{OSIRP:LRT} needs to be clarified. Under a normality assumption on the
errors, $2\,\log \lambda_I$ is, indeed, a proper likelihood ratio statistic; otherwise, it is more accurately a \emph{residual sum of 
squares statistic} which can be interpreted as a `working likelihood ratio statistic' where the normal likelihood is used as a working 
likelihood. In this paper, we do not assume normality of errors but continue to use the term LR statistic for $2\,\log \lambda_I$ in the 
above sense. 
\end{TangRem}

\subsection{Adaptive Two-Stage Procedures}\label{sec:2stage}
As noted in Introduction, adaptive two stage procedures can lead to
accelerated convergence rates and hence to sharper confidence
intervals for $d_0$. The main steps of such a two-stage fully nonparametric procedure are outlined next:
\begin{enumerate}
\item
Denote by $p\in (0,1)$ the sample proportion to be allocated in the first stage
and  by $n_{1}=\lfloor np\rfloor$ and $n_{2} = n - n_{1}$, the corresponding first and second
stage sample sizes, respectively.
\item
Generate the first stage data $\{(X_{1,i}, Y_{1,i})\}_{i=1}^{n_{1}}$
with a design density $g_{1}$ on $[a,b]$.
Then, compute a first stage monotone non-parametric estimator $\hat m_1$ of $m$ 
and obtain the corresponding first stage estimator 
$d_{1,I} = \hat m^{-1}_{1}(\theta_0)$ of $d_{0}$
for a prespecified value $\theta_{0}$.
\item
Specify the second stage sampling interval
$[L_{1}, U_{1}] = [ d_{1,I} \pm C_{1}n_{1}^{-\gamma_{1}}] \cap [a,b]$
where $C_{1}>0$ and
$0<\gamma_{1}<\gamma^*<1/2$, 
$\gamma^{\star}$ being the convergence rate of $\hat{d}_{1}$.
\item
Obtain the second stage data $\{(X_{2,i},Y_{2,i})\}_{i=1}^{n_{2}}$
with a design density $g_{2}$ on $[L_{1},U_{1}]$.
Employ these data and a non-parametric procedure (which could be different from the one used previously)
to compute a monotone second stage estimator $m_{2,I}$ and, as in the first stage, the corresponding $d_{2,I}$.
\item
Construct confidence intervals for $d_{0}$ using the asymptotic distribution of $d_{2,I}$.
\end{enumerate}

\begin{TangRem}
Choosing $\gamma_1 < \gamma^{\star}$ ensures that the stage two sampling interval contains $d_{0}$ with probability going to 1.
\end{TangRem}

\subsection{Asymptotic Properties of Two-Stage Estimators}

We discuss the properties of the two-stage procedure, where isotonic regression is employed in both stages (henceforth, IR+IR).

\begin{TangProp}\label{TSIRP:Prop:d2I:Wald}
    Consider the IR + IR procedure. Let the design density at stage two be given by:
    $g_2(x) = (C_1\,n_1^{-\gamma_1})^{-1}\,\psi((x - d_{1,I})/C_1\,n_1^{-\gamma_1})$ where $\psi$ is
    a Lebesgue density on $[-1,1]$ that is positive at 0 and continuous in a neighborhood of 0.  Thus, $g_2$
    is simply $\psi$ renormalized to the sampling interval at stage two. Assume that $m'$, the derivative of $m$, exists and
    is continuous in a neighborhood of $d_0$ and $m^{'}(d_{0}) > 0$. Let $d_{2,I} = m_{2,I}^{-1}(\theta_0)$ where $m_{2,I}$ is the isotonic
    estimator of $m$ constructed from the second stage data. Then, $n^{(1+\gamma_{1})/3}(d_{2,I} - d_{0})
        \WC
        C_{d_{2,I}}
        \CherDist$,
where
$C_{d_{2,I}}= C_{d_{I}}\left(\frac{C_{1}}{(1-p)p^{\gamma_{1}}\psi(0)}\right)^{1/3}$.
\end{TangProp}

From Proposition \ref{TSIRP:Prop:d2I:Wald}, a Wald-type $1-\alpha$ asymptotic confidence interval for $d_{0}$ is
given by
\begin{equation}\label{TSIRP:Prop:d2I:Wald:CI}
[d_{2,I}\pm n^{-(1+\gamma_{1})/3} \widehat{C_{d_{2,I}}}
q(\mathcal{Z}, 1-\alpha/2)].
\end{equation}

\begin{TangRem}
A consequence of the accelerated rate of convergence obtained with the IR+IR strategy is that
the asymptotic relative efficiency (ARE) of the two-stage estimator $d_{2,I}$ with respect to the
one-stage estimator $d_{I}$ is
\begin{equation*}
    ARE(d_{2,I},d_{I})
    = \frac{s.d.(d_{I})}{s.d.(d_{2,I})}
    = \kls \frac{(1-p)p^{\gamma_{1}}\psi(0)}{C_{1}g(d_{0})} \krs^{1/3}n^{\gamma_{1}/3}
    \rightarrow \infty
    \text{ as } n\rightarrow \infty.
\end{equation*}
\end{TangRem}

Note that, in the generic description of the two-stage procedure above, we use a confidence interval for $d_{0}$ that relies on the asymptotic
distribution of a point estimate computed at stage two. However, this is not the only way to proceed at stage two. Having collected the second stage data at the beginning of Step 4, we can bypass point estimation altogether and construct a confidence interval using likelihood ratio inversion. This alternative possibility is discussed below. Also, as will be explained in the practical implementation, the construction of $[L_1, U_1]$ is achieved in practice by constructing a high probability confidence interval for $d_{0}$
from the stage one data. This also opens up the possibility of bypassing point estimates at stage one in favor of a likelihood ratio inversion based confidence interval, a point that we come to later. 
\newline
\newline
An alternative LR-type CI can be constructed as follows:
the LR-type test statistic at stage two for testing $H_0: d_0 = m^{-1}(\theta_0)$  is
\begin{equation}\label{TSIRP:LRT}
            2\log\lambda_{2,I}
            = 2\log\lambda_{2,I}(d_0)
            = 2 \klm l_{n}(m_{2,I}, \hat{\sigma})-l_{n}(m_{2,Ic}, \hat{\sigma}) \krm,
\end{equation}
where $l_{n}(m, \sigma)=-\frac{1}{2\sigma^{2}}\sum_{i=1}^{n_{2}}(Y_{2,i}-m(X_{2,i}))^{2}$, $m_{2,I_c}$ is the
constrained estimator of $m$ under the null hypothesis $H_{0}$ and 
$\hat{\sigma}$ is a consistent estimate of $\sigma$. 

\begin{TangProp}\label{TSIRP:Prop:d0:LRT}
Under the assumptions of Proposition  \ref{TSIRP:Prop:d2I:Wald}, and the null hypothesis
$H_{0}$: $m^{-1}(\theta_{0})=d_{0}$ holding true, we have
$2\log\lambda_{2,I} \WC \mathbb{D}$,
where $\mathbb{D}$ is as before. 
\end{TangProp}

Finally, from Proposition \ref{TSIRP:Prop:d0:LRT},
an LR-type $(1-\alpha)$ asymptotic confidence interval for $d_{0}$
is given by
\begin{equation}\label{TSIRP:Prop:d0:LRT:CI}
\{x\in[a,b]: 2\log\lambda_{2,I}(x)\leq q(\mathbb{D}, 1-\alpha)\}.
\end{equation}
For the theoretical derivations in connection with Propositions \ref{TSIRP:Prop:d2I:Wald} and \ref{TSIRP:Prop:d0:LRT}, see the Appendix.

\begin{TangRem}
We have focused on the case of two-stage adaptive designs and the acceleration of the convergence rate by an IR+IR strategy. 
Obviously, one can extend it to multiple stages and continue using isotonic regression. As outlined in Section S1 in the
Supplement, it can be established that the convergence rate of such a procedure would come {\em arbitrarily close} to the 
$\sqrt{n}$ parametric rate, if enough stages are employed, {\em but would not achieve it}.
\end{TangRem}

\subsection{Implementation Issues}

We discuss, next, the main steps for implementing the IR+IR strategy in practice.
Specifically, we address the following: (i) estimation of $\sigma^2$, (ii) estimation of $m'$, (iii)
determination of second stage sampling interval $[L_1,U_1]$, (iv) the first stage sampling proportion $p$. 
\newline
{\bf Implementation of IR + IR:} For the estimation of $\sigma^2$ at Stage 1, we employ the nonparametric estimator proposed by
\cite{Gasser1986}, and for the estimation of $m'(d_0)$, the local quadratic regression procedure
proposed by \cite{Fan1996}; some details are provided in Section 3. Next comes the determination of the 
second stage sampling interval. Recall, that the theoretical formula for the interval is given by $ [d_{1,I} \pm C_{1}n_{1}^{-\gamma_{1}}]$, with $C_{1}>0$ and $\gamma_{1}\in(0,1/3)$. While any such interval will contain $d_0$ with probability going to 1 in the long run, in practice we would like to ensure that our prescribed sampling interval $[L_1,U_1]$ does trap $d_0$ with high probability. The practical determination of $[L_1,U_1]$ is therefore 
achieved through a high probability confidence interval for $d_0$ from Stage 1 data.  Consider the the following $1-\beta$ Wald-type confidence interval
\begin{equation}\label{PTSIRP:L1U1}
    [d_{1,I} \pm
    n_{1}^{-1/3} \hat C_{d_{I}}g_{1}(d_{1,I})^{-1/3}
    q(\mathcal{Z},1-\beta/2)]\cap [a,b],
\end{equation}
where the computation of
$\hat C_{d_{I}}$ involves estimating both $\sigma^{2}$ and $m'(d_{0})$ and where
$\beta$ is a small positive number such as 0.01. Using this, in practice, as $[L_1,U_1]$ amounts to
choosing $C_{1}$ and $\gamma_{1}$ such that
$C_{1}n_{1}^{-\gamma_{1}}
    = n_{1}^{-1/3} \hat C_{d_{I}}g_{1}(d_{1,I})^{-1/3}q(\mathcal{Z},1-\beta/2)$.
That is,
    $\gamma_{1} = 1/3
    ~~\text{and}~~
    C_{1}
    = \hat C_{d_{I}}g_{1}(d_{1,I})^{-1/3}q(\mathcal{Z},1-\beta/2)$.
Although $\gamma_{1}=1/3$ is not in $(0, 1/3)$ as required by our theoretical results, it nevertheless provides a
good approximation in practice, since it lies at the boundary of that interval. 

As far as the first stage sampling proportion is concerned, one would like to choose this in such a way as to increase the precision of the second stage 
isotonic estimator. Proposition \ref{TSIRP:Prop:d2I:Wald} shows that the second stage estimator is asymptotically unbiased and that its standard deviation is proportional to $\{(1-p)\,p^{\gamma_1}\}^{-1/3}$. For a fixed $\gamma_1$, this is minimized when $\log (1-p) + \gamma_1\,\log p$ is maximized, which 
happens when $p = p_{opt} = \gamma_1/(1+\gamma_1)$. As $\gamma_1$ approaches $1/3$, $p_{opt}$ approaches $1/4$. Thus, 
the optimal practical allocation of budget at Stage 1 is 25\%.

From Stage 2 data, we can construct a confidence interval for $d_{0}$ based on $d_{2,I}$ following Proposition 
\ref{TSIRP:Prop:d2I:Wald} in which case both $m'(d_{0})$ and $\sigma^2$ need to be updated.  Alternatively, we 
can use likelihood ratio inversion to get a CI of desired coverage for $d_{0}$, following Proposition \ref{TSIRP:Prop:d0:LRT} in which case only the estimate of $\sigma^2$ needs to be updated. Finally, a third option is to bypass the estimation of $m'(d_{0})$ altogether by 
prescribing a high probability LR based confidence interval for $d_{0}$ as $[L_1, U_1]$ in Stage 1 and then using LR inversion at Stage 2 as well. While this procedure does not quite fall within the purview of our theoretical results it is a natural methodological choice; furthermore, comparisons among these three approaches based on elaborate simulation studies demonstrate that it is superior to the other two in practice.

\section{Performance Evaluation of the Adaptive Procedures}
\label{sec3}

\begin{figure}
  \begin{center}
  \includegraphics[scale=0.38]{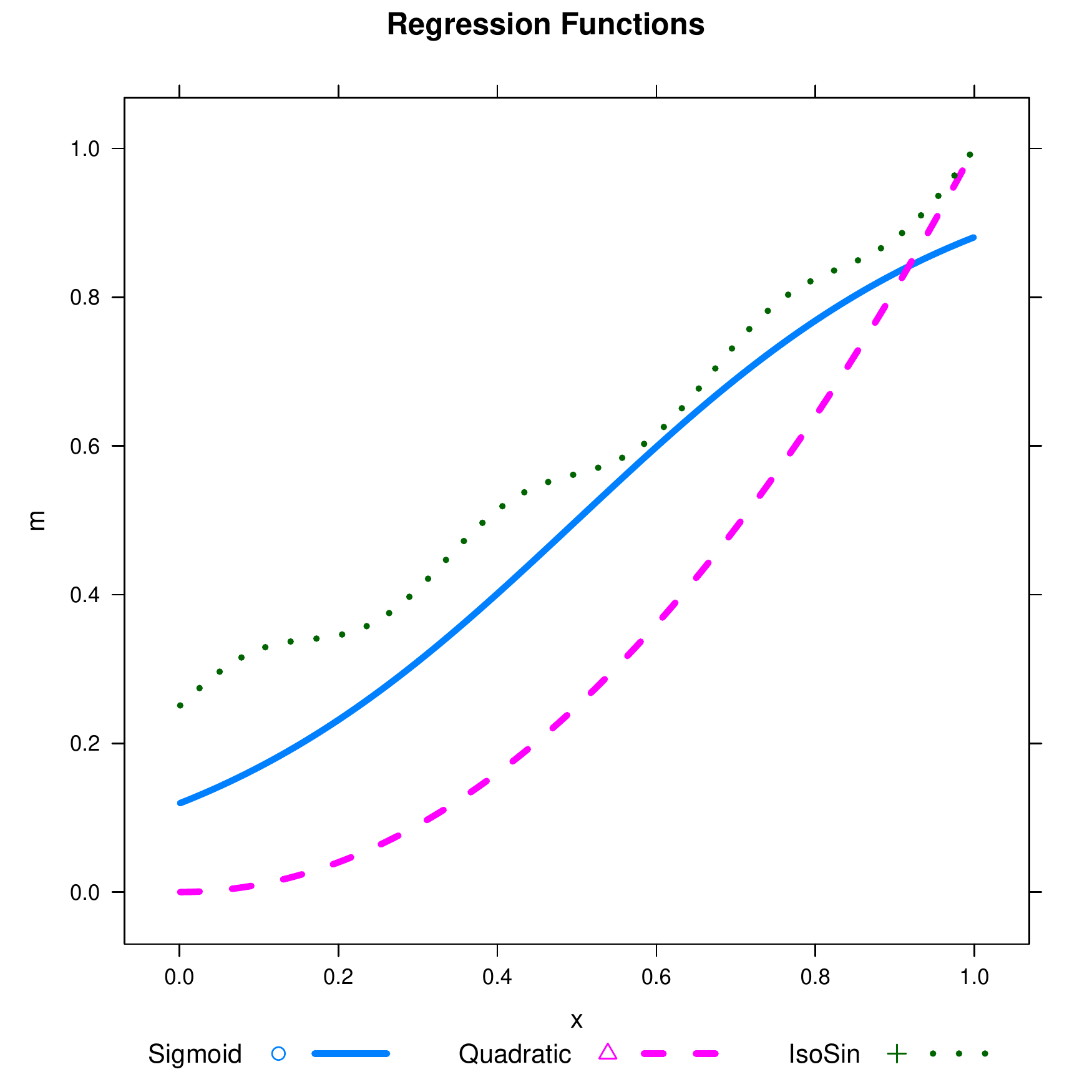}
  \includegraphics[scale=0.38]{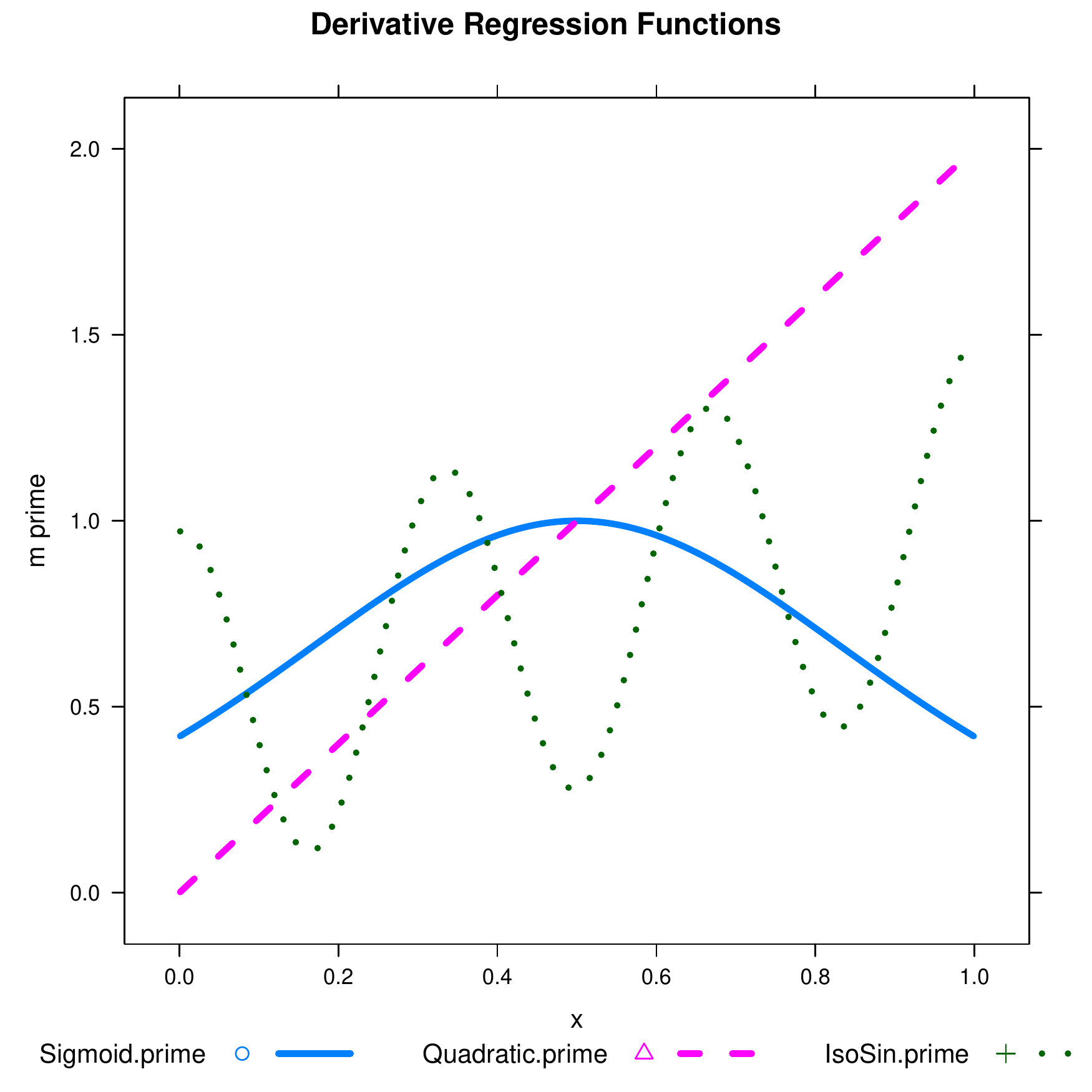}\\
  \caption
  [Regression functions and their derivatives]
  {The left plot shows the regression functions:
  sigmoid, quadratic and isotonic sine functions.
  The right plot shows their derivatives.}
  \label{paper:fg:regressionFunctions:derivative}
  \end{center}
\end{figure}

In this study, the following procedures are compared: (i) practical one-stage
procedure based on isotonic regression (POSIRP) with Wald and LR CIs, (ii) practical two-stage
procedure based on isotonic regression (PTSIRP-Wald) for both stages and using Wald CIs both
for selecting $(L_1,U_1)$ and constructing the final CI, (iii) practical two-stage procedure (PTSIRP-LR) similar to (ii)
but employing LR CIs in both stages and (iv) the procedure from \cite{tang2011} that uses isotonic regression 
followed by a local linear approximation and bootstrapping for constructing CIs for $d_0$ (PABLTSP). 
The use of the qualifier `Practical' before the 
various procedures above is to emphasize the point that they involve estimates of nuisance parameters, as explained
below.

The simulation settings are as follows: the design space is the $[0,1]$ interval and the
regression functions considered: (i) the sigmoid function
 $m(x) = \exp(4(x-0.5))/[1+\exp(4(x-0.5))]$,  the quadratic function
 $m(x) = x^{2}$ and the isotonic sine function
 $m(x) = (1/40)\sin(6\pi x) + 1/4 + (1/2)x + (1/4)x^{2}$.
The target point $d_{0}$ is 0.4, 0.5 or 0.6, while the random error follows a $N(0, \sigma^{2})$ distribution,
 with $\sigma$ taking values 0.1 and 0.3. The total sample size $n$ ranges from 100 to 500 in increments of 100.
All design densities $g$, $g_{1}$ and $g_{2}$ are uniform, while the confidence level for all CIs 
is set to $0.95$. The results presented are based on 1000 replicates. For PABLTSP, we set the first stage sample proportions $p=0.7$ in order obtain accurate coverage rates for all functions. (Yet, as shown in Figure \ref{illustration1}, in some cases good coverage rates are achieved at the cost of large average lengths.) For all other two stage procedures, we set $p$ to be the asymptotically optimal proportion of 0.25. 
The quantiles of $\DD$ and $\ZZ$ for
constructing the second-stage sampling intervals for PTSIRP 
are set to be $4$ and $2$, respectively, corresponding to $\beta = 0.01$. 

When estimating $\sigma$ and $m'(d_0)$ in the second stage, only Stage 2 points are used in order to stay strictly within the scope of the methods used for these purposes. With smaller budgets as in the real data example, we follow the natural practice of combining both stage samples for updating estimates of $\sigma$ and $m'(d_0)$, which makes second stage results more reliable.

To gain insight into the simulation results, we depict the plots of the functions under consideration together
with their derivatives (see  Figure \ref{paper:fg:regressionFunctions:derivative}).

\begin{figure}
  \begin{center}
  \includegraphics[width=.49\columnwidth, trim=0.1cm 0.4cm 0.4cm 0.5cm, clip=true]{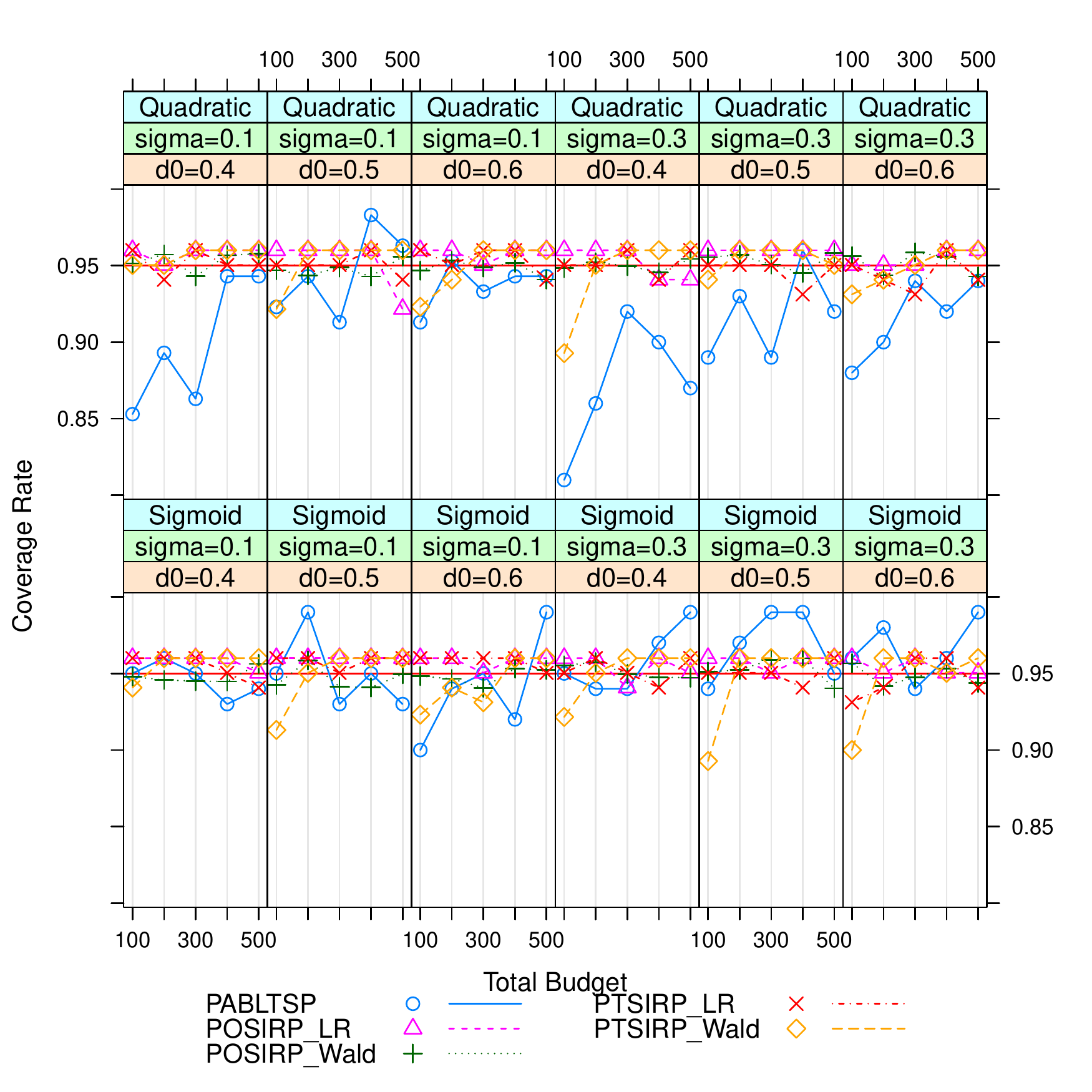}
  \includegraphics[width=.49\columnwidth, trim=0.1cm 0.4cm 0.4cm 0.5cm, clip=true]{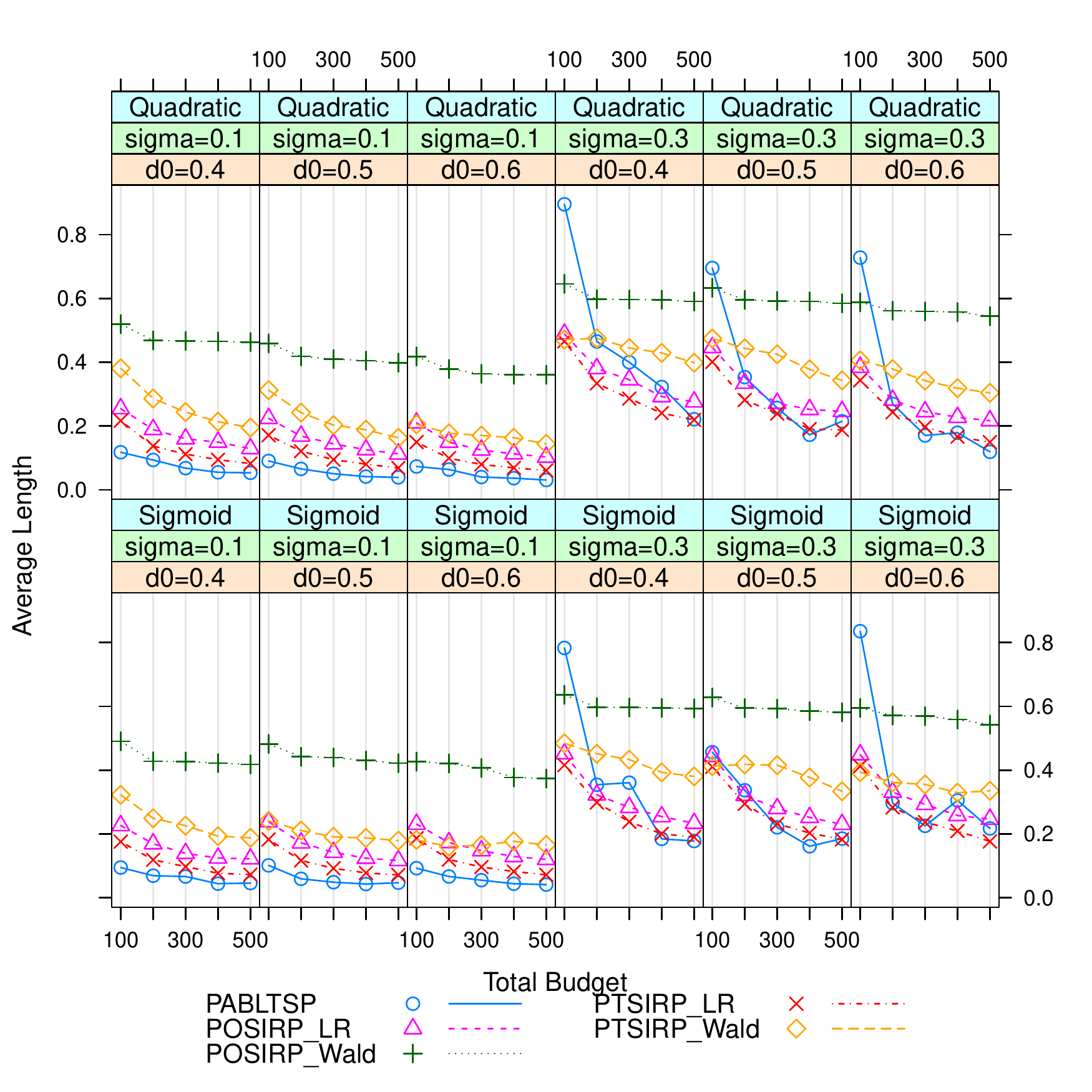}\\
  \caption
  [Coverage rates and average lengths with the sigmoid and quadratic functions]
  {The left and right panels show the coverage rates and average lengths
of the 95\% confidence intervals for $d_{0}$ from the practical procedures
with the sigmoid and quadratic functions
and different values of $\sigma$, $d_{0}$ and $n$.}
  \label{paper:FG:SigmoidAndQuadratic:CRAndAL2}
  \end{center}
\end{figure}

\begin{figure}[!h]
  \begin{center}
  \includegraphics[width=.49\columnwidth, trim=0.1cm 0.4cm 0.4cm 0.5cm, clip=true]{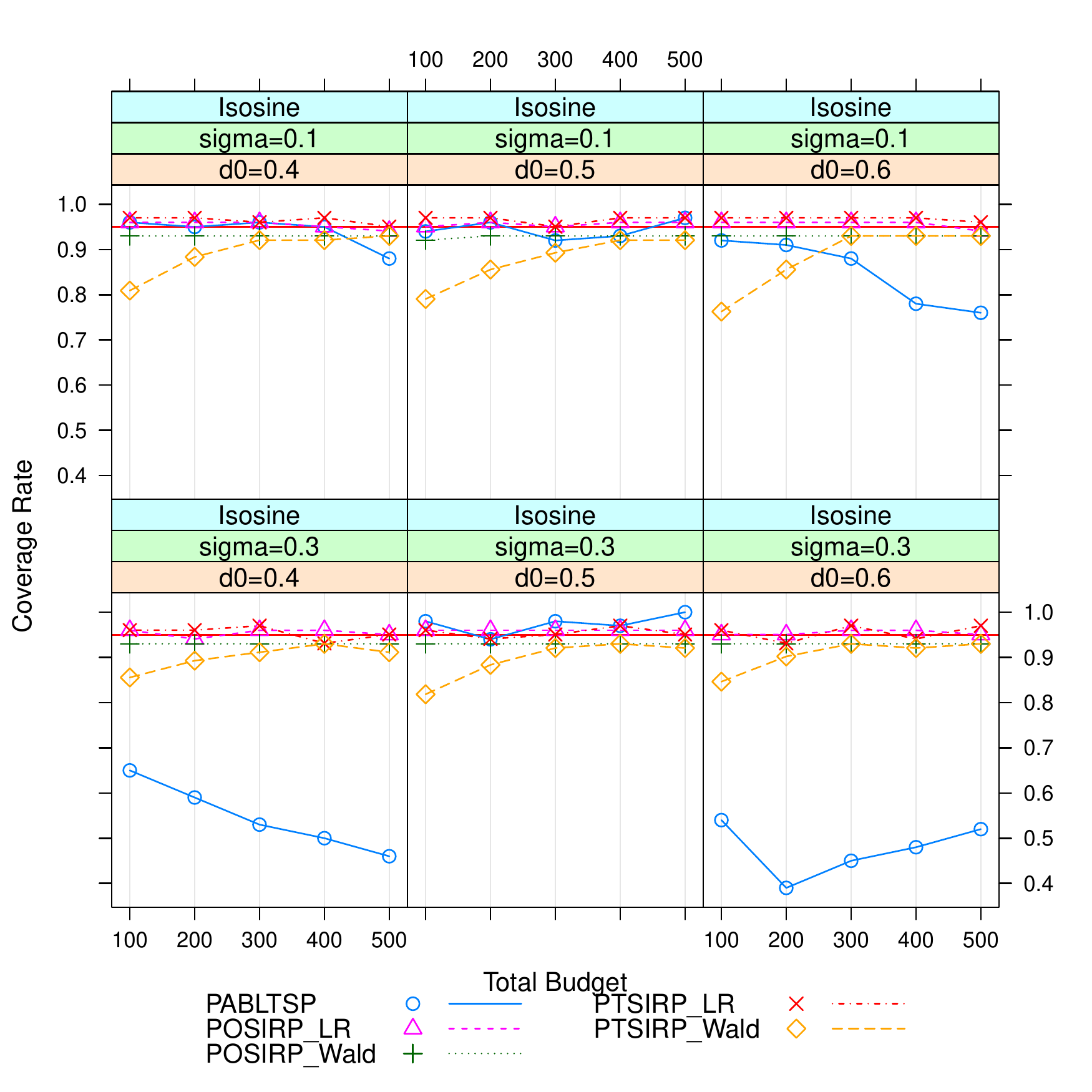}
  \includegraphics[width=.49\columnwidth, trim=0.1cm 0.4cm 0.4cm 0.5cm, clip=true]{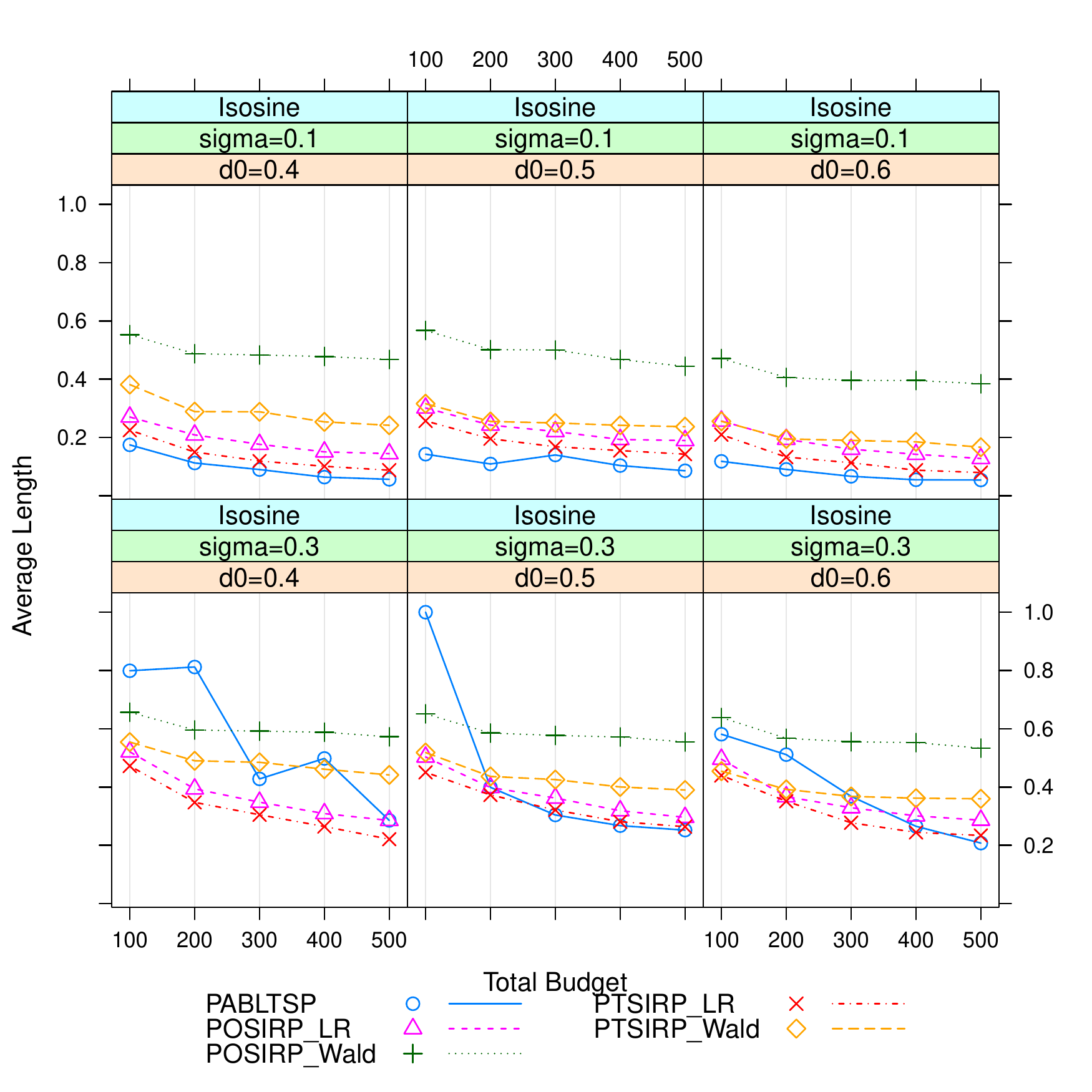}\\
  \caption
  [Coverage rates and average lengths with the isotonic sine function]
  {The left panel shows the coverage rates
of the 95\% confidence intervals for $d_{0}$ from the practical procedures
with the isotonic sine functions
and different values of $\sigma$, $d_{0}$ and $n$.
The right panel shows the average lengths
of the 95\% confidence intervals for $d_{0}$.}
  \label{paper:FG:isotonic sine:CRAndAL2}
  \end{center}
\end{figure}

The coverage rates and average lengths of the $95\%$ confidence intervals for $d_{0}$
are shown in Figures \ref{paper:FG:SigmoidAndQuadratic:CRAndAL2} and \ref{paper:FG:isotonic sine:CRAndAL2}.

It can be seen that for the quadratic and sigmoid functions, the proposed two-stage procedures perform well with
the coverage being about the nominal level $95\%$ for all $d_0$'s,
sample sizes and noise levels considered. Further, their average lengths are fairly comparable.
In contrast, PABLTSP shows inferior performance for larger noise and smaller sample sizes for the quadratic function.

The isotonic sine function proves the most challenging. The left panel of 
Figure \ref{paper:FG:isotonic sine:CRAndAL2}
shows the coverage rates
of the practical procedures. As discussed in the introduction and seen in the figure, 
this function exhibits strong nonlinearity causing the 
the local linear approximation PABLTSP to feature very poor coverage rates. 
Also, note that, for the case with $d_{0}=0.5$
the coverage rates of the confidence intervals from
POSIRP-Wald and PTSIRP-Wald  
are consistently lower than $95\%$.
This behavior is caused by inaccurate estimation of $m'(d_{0})$ as illustrated in the Supplementary material (Section S2, Figure 1). The true value of $m'(d_{0})$ is around 0.279
and the corresponding kernel estimators of $m'(d_{0})$ are usually
around 0.75, significantly larger than the true value. 
This makes the confidence interval far too short to cover $d_{0}$
and consequently, the coverage rates behave erratically. 

Full details for the estimation of $m'(d_{0})$, which utilizes a local quadratic regression procedure, are available 
in Section 4 of \cite{tang2011}. An asymptotically optimal bandwidth, given in 
equation (3.20) on page 67 of \cite{Fan1996}, is employed for this purpose. This local bandwidth minimizes 
the asymptotic MSE, and indeed with large sample sizes, we find that $m'(d_{0})$ 
is estimated accurately and the coverage rates approach the nominal level. Further emphasizing 
the importance of the derivative estimate and as illustrated in Figure 2 of the Supplementary material, if we 
repeat the procedures with perfect knowledge of 
$m'(d_0)$, then coverage rates are about the nominal level of $95\%$ 
for the sample sizes considered.

Fortunately, for this wiggly isotonic sine function,
POSIRP-LR and PTSIRP-LR
have good coverage rates
for all simulation cases. This indicates that LR-type confidence intervals
are usually robust with different regression functions.
The average lengths of the confidence intervals
are shown in the right panel of Figure \ref{paper:FG:isotonic sine:CRAndAL2}.
Unsurprisingly, PTSIRP-LR achieves shorter average lengths since it is a two-stage procedure.

In summary, we find that when the underlying regression function is well-behaved,
the more aggressive PABLTSP performs well. However, the conservative but stable PTSIRP-LR
offers a robust procedure that performs well, even when the underlying function 
exhibits strong nonlinearities.

\section{An Application to Fuel Efficiency Standards}
\label{TSPInvFun:paper:DataAnalysis}

As discussed in Introduction, car fuel efficiency (FE) is an important issue for both manufacturers and
consumers, due to new CAFE standards. Note that while the CAFE standards are regulated by the NHTSA, the vehicle 
 FE is assessed by the Environmental Protection Agency (EPA). From 2008 onwards, the EPA 
measures the fuel efficiency of a vehicle in two testing modes: 
city and highway, taking into consideration different speeds and acceleration, as well as air conditioning usage
and colder outside temperatures, in an effort to better approximate real-world fuel efficiency.
From the unadjusted city and highway fuel efficiency,
the unadjusted combined fuel efficiency is calculated as follows (see {\tt www.epa.gov}):
\begin{equation*}
    \text{Combined FE} =
    \frac{1}{.495/\text{City FE} + .351/\text{Highway FE}} + .15.
\end{equation*}

The data for this study were extracted from the government website 
{\tt www.fueleconomy.gov} that includes all FE data for all 2012 car models available to
US consumers. This data set contains the unadjusted city, highway and combined fuel efficiency
for 3979 models, together with their horse power. 
We collected the horse power data for 1477 non-hybrid vehicles with automatic transmission gearboxes and
natural aspiration engines (i.e. excluding turbo engines and plug-in hybrid vehicles),
in order to have a relatively homogeneous data set.

The objective of our analysis is to estimate the following model $FE=m(HP)$ (or $HP=m^{-1}(FE)$ and then
identify the horse power
at which the combined FE is equal to 30 MPG, around the 2011 CAFE standard. Hence, we are interested in
estimating $d_0 = m^{-1}(30)$.

The scatter plot in the left panel of
Figure \ref{paper:FG:DataAnalysis:HorsePowerCombinedFuelEfficiencyAirAspiration}
shows the combined FE of these 1477 vehicles as a function of their horse power and
indicates a decreasing relationship. Notice that there are multiple vehicle models with the same
horse power, but different FE. To simplify the analysis, we add a small jitter to the original  
horse power to obtain a unique horsepower for every FE observation, whose scatterplot
is given in the right panel of  Figure \ref{paper:FG:DataAnalysis:HorsePowerCombinedFuelEfficiencyAirAspiration}. 
The jitter added is between $\pm1$ to ensure that the ordering of samples by horsepower remains unchanged.

Given that this is an 'observed' data set, we will {\em emulate} the design setting (for a similar strategy see
also \cite{lan2009}) for a budget of size 80. Both one stage and two stage procedures will be examined.
For one stage procedures, 80 horse powers equally spaced are originally selected and the closest ones
in the data constitute the final covariate values, together with the corresponding responses. 
For two stage procedures, we select a portion $p=0.5$ in the first stage and hence select
40 horse powers in the first stage as previously described. After obtaining the second stage sampling interval
$(L_1,U_1)$ we choose with an analogous strategy the remaining 40 points. (Given the relatively modest budget, we chose not 
to use the asymptotically optimal allocation of $25\% + 75\%$.) 

Finally, the ``true" value of $d_0$ is obtained by using isotonic regression on the entire sample of 1477 observations 
and is estimated to be around 187.

\begin{figure}[!h]
  \begin{center}
  \includegraphics[width=.45\columnwidth,trim=0.4cm 0.85cm 0.4cm 1.45cm, clip=true]{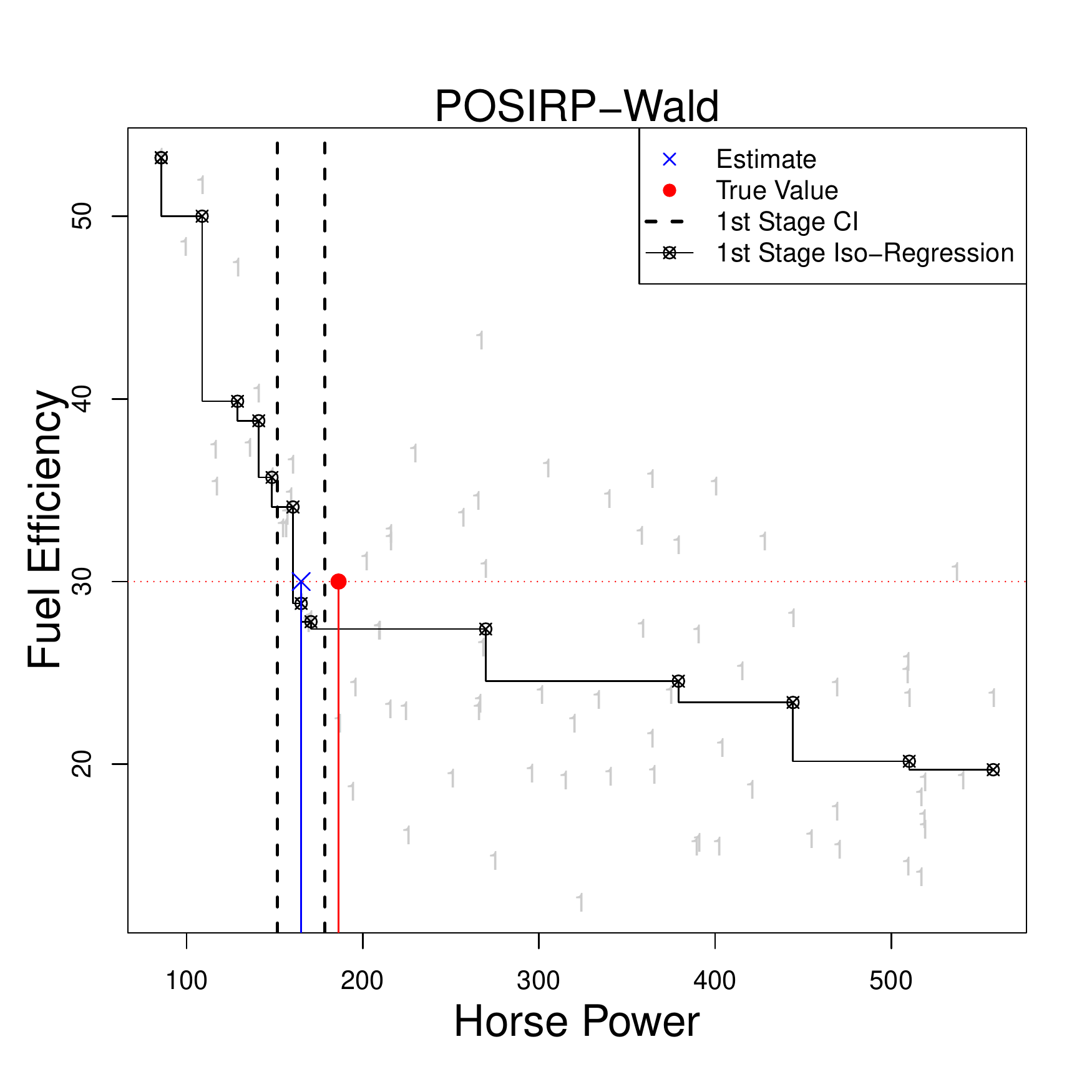}  
  \includegraphics[width=.45\columnwidth,trim=0.4cm 0.85cm 0.4cm 1.45cm, clip=true]{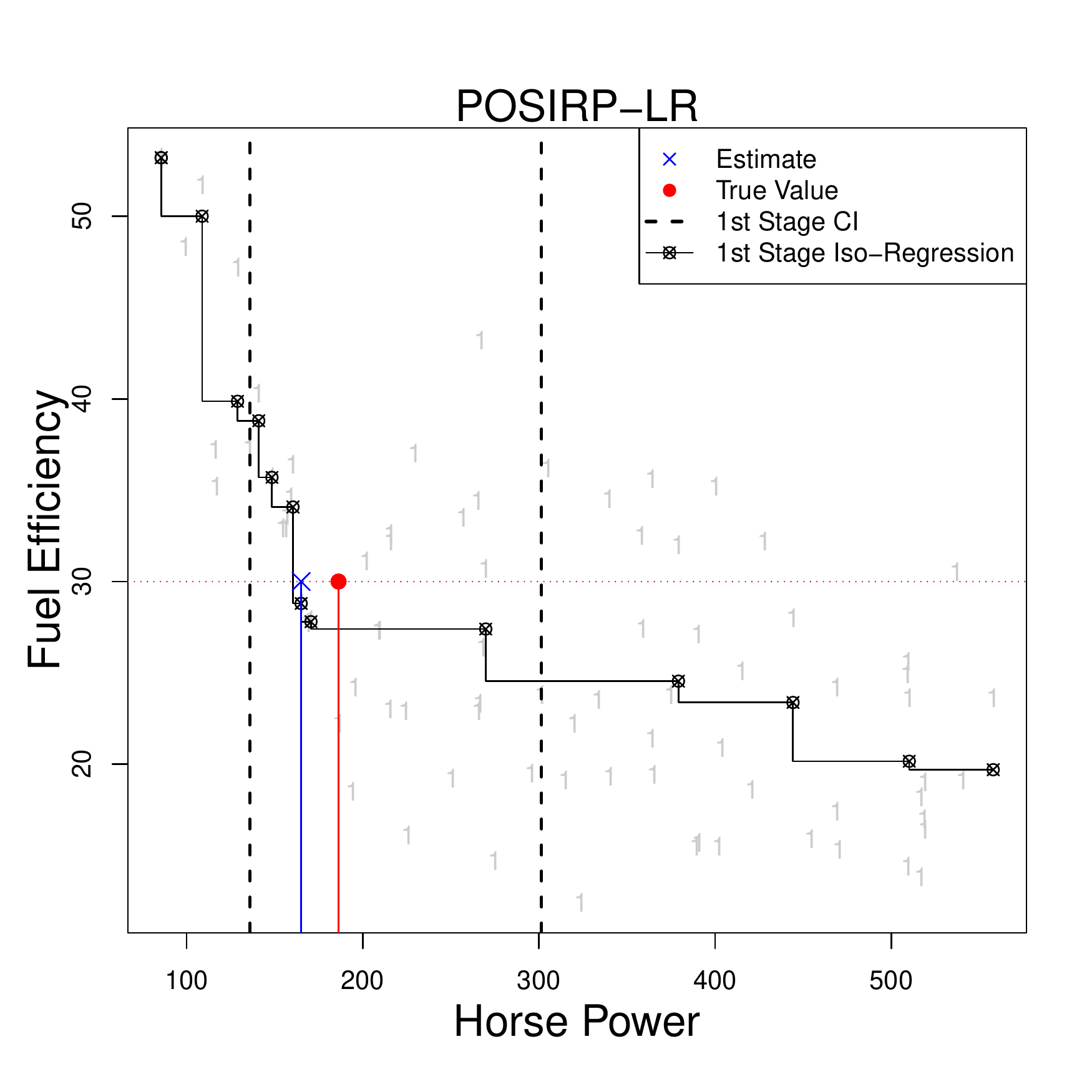} \\
    \includegraphics[width=.32\columnwidth,trim=0.4cm 0.85cm 0.4cm 1.45cm, clip=true]{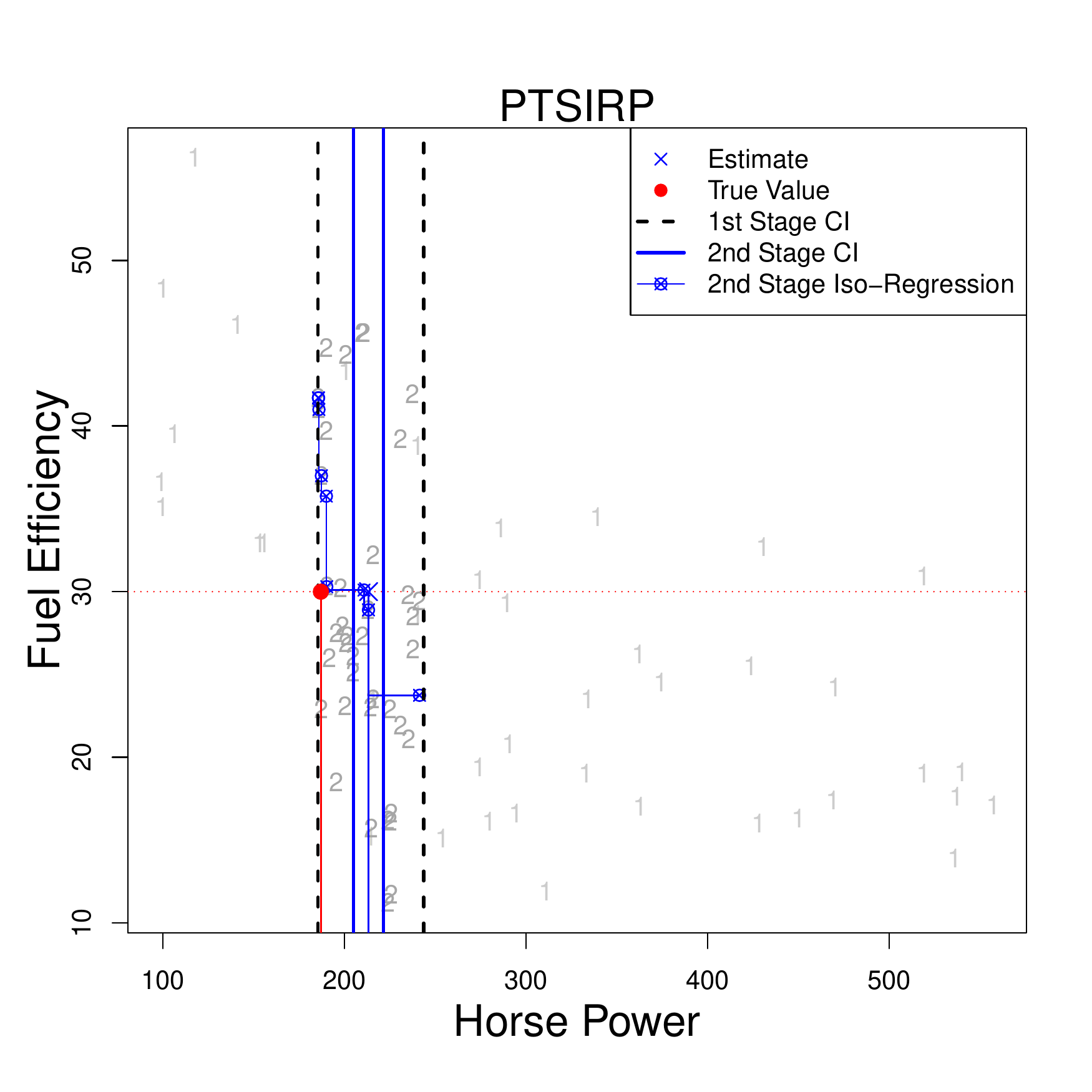}
  \includegraphics[width=.32\columnwidth,trim=0.4cm 0.85cm 0.4cm 1.45cm, clip=true]{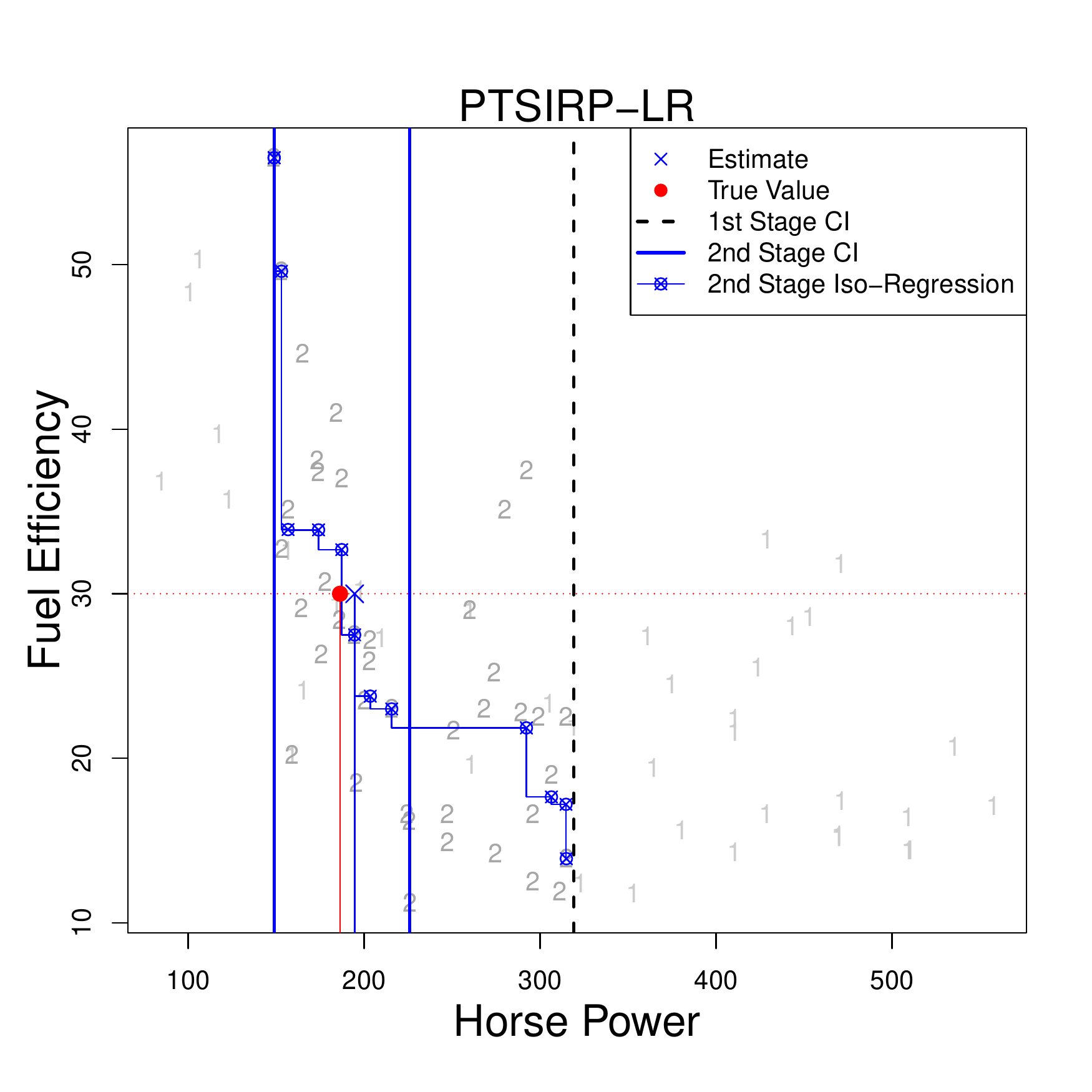}
  \includegraphics[width=.32\columnwidth,trim=0.4cm 0.85cm 0.4cm 1.45cm, clip=true]{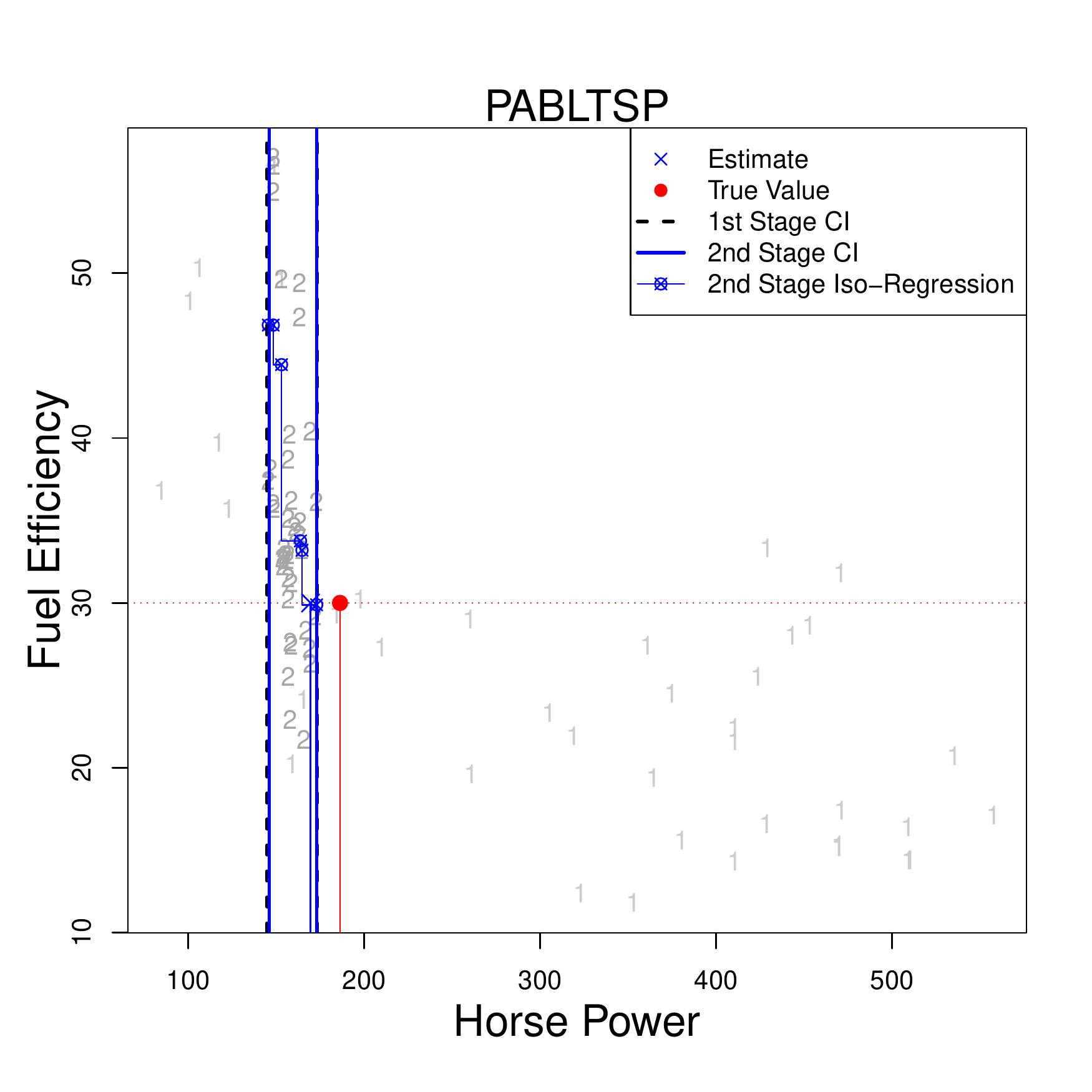}
  \caption
  [Plots for the data analysis results]
  {The top panels show one stage procedures: POSIRP-Wald and POSIRP-LR.
   The bottom panels show two stage procedures: PTSIRP, PTSIRP-LR and PABLTSP.
   Numbers denote first and second stage samples, vertical lines denote corresponding confidence intervals, and `X' marks the final point estimate.}
  \label{paper:FG:DataAnalysis:PracticalProcedures}
  \end{center}
\end{figure}

\begin{table} \small
  \centering
  \caption[Data analysis results for the five practical procedures]
  {Data Analysis Results for the Five Practical Procedures}
  \label{paper:Table:DataAnalysis:PracticalProcedures}
\begin{tabular}{|c|c|c|c|c|c|c|}
  \hline
  Procedure & Estimator & ``Bias" & 95\% CI & Coverage & Length & $n$ \\ \hline \hline            
  POSIRP-Wald & 165.022 & 21.978 & $[151.595, 178.450]$ & No & 26.855 & 80 \\                
  POSIRP-LR & 165.022 & 21.978 & $[135.887, 301.439]$ & Yes & 165.552 & 80 \\            
  PTSIRP-Wald & 213.221 & 26.221 & $[205.042, 221.400]$ & No & 16.358 & 40,40 \\
  PTSIRP-LR & 194.557 & 7.557 & $[148.812, 225.955]$ & Yes & 77.143 & 40,40 \\      
  PABLTSP & 169.509 & 17.491 & $[145.982, 172.964]$ & No & 26.982 & 40,40 \\
  \hline
\end{tabular}
\end{table}
%
%
The five procedures considered are: POSIRP-Wald, POSIRP-LR, PTSIRP-Wald, PTSIRP-LR and PABLTSP. The fitted models are shown in Figure \ref{paper:FG:DataAnalysis:PracticalProcedures}
and the confidence intervals obtained summarized in Table \ref{paper:Table:DataAnalysis:PracticalProcedures}.
It is interesting to note that only the LR based procedures produce confidence intervals that cover the ``true value.'' The two-stage procedure PTSIRP-LR achieves much shorter interval lengths. The PTSIRP-Wald CIs are too short, resulting in their missing the ``true value."

Additional results from utilizing different allocations in the two stages are provided in Figure 3 of the 
Supplementary material, where we see that in almost all cases PTSIRP-LR tends to cover the 
``true'' value of $187$ with better point estimates. 
POSIRP-Wald and PTSIRP-Wald continue to struggle due to estimation difficulties with $m'(d_0)$. 

\begin{figure}
\begin{center}
\includegraphics[width=0.48\columnwidth]{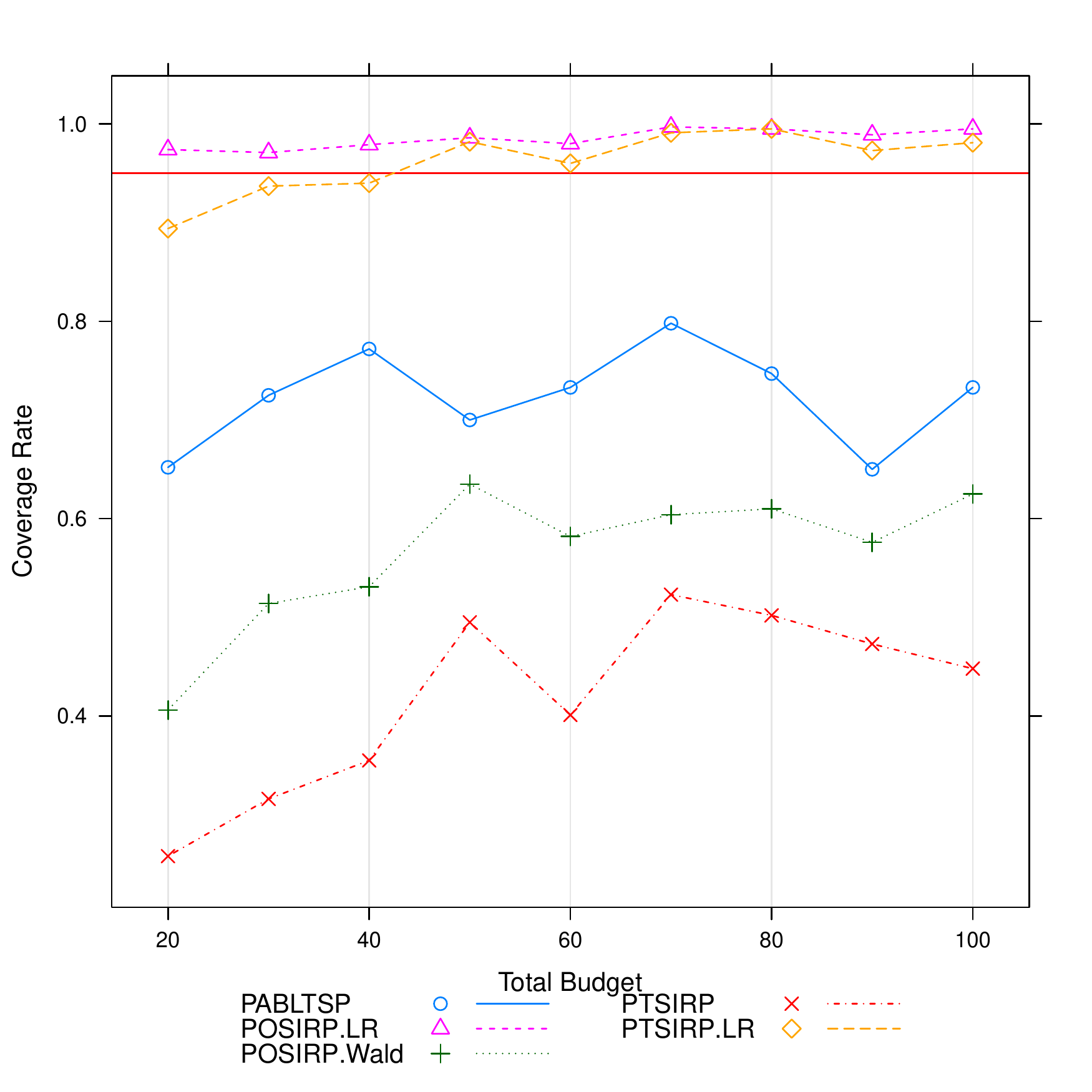}
\includegraphics[width=0.48\columnwidth]{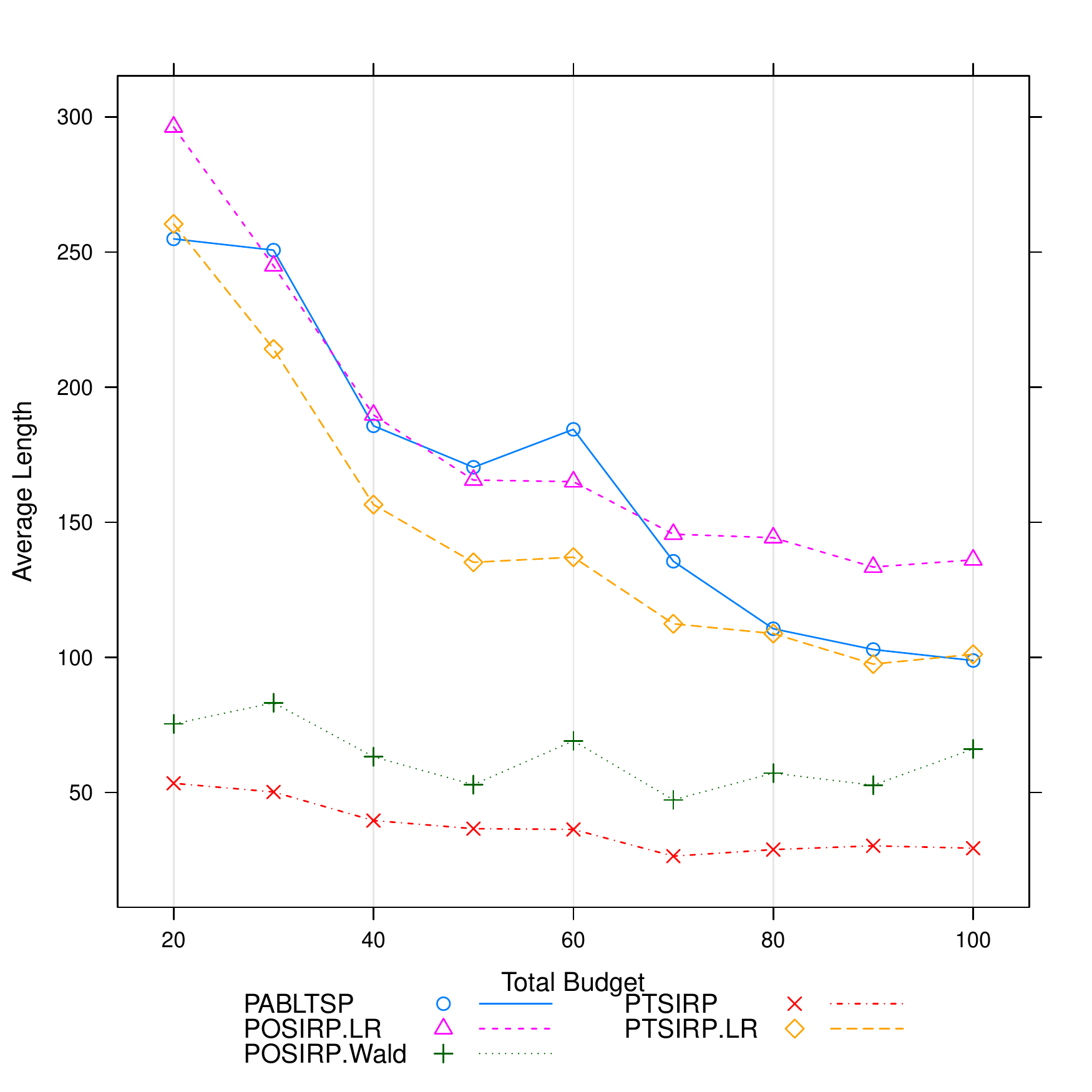}\\
\includegraphics[width=0.48\columnwidth]{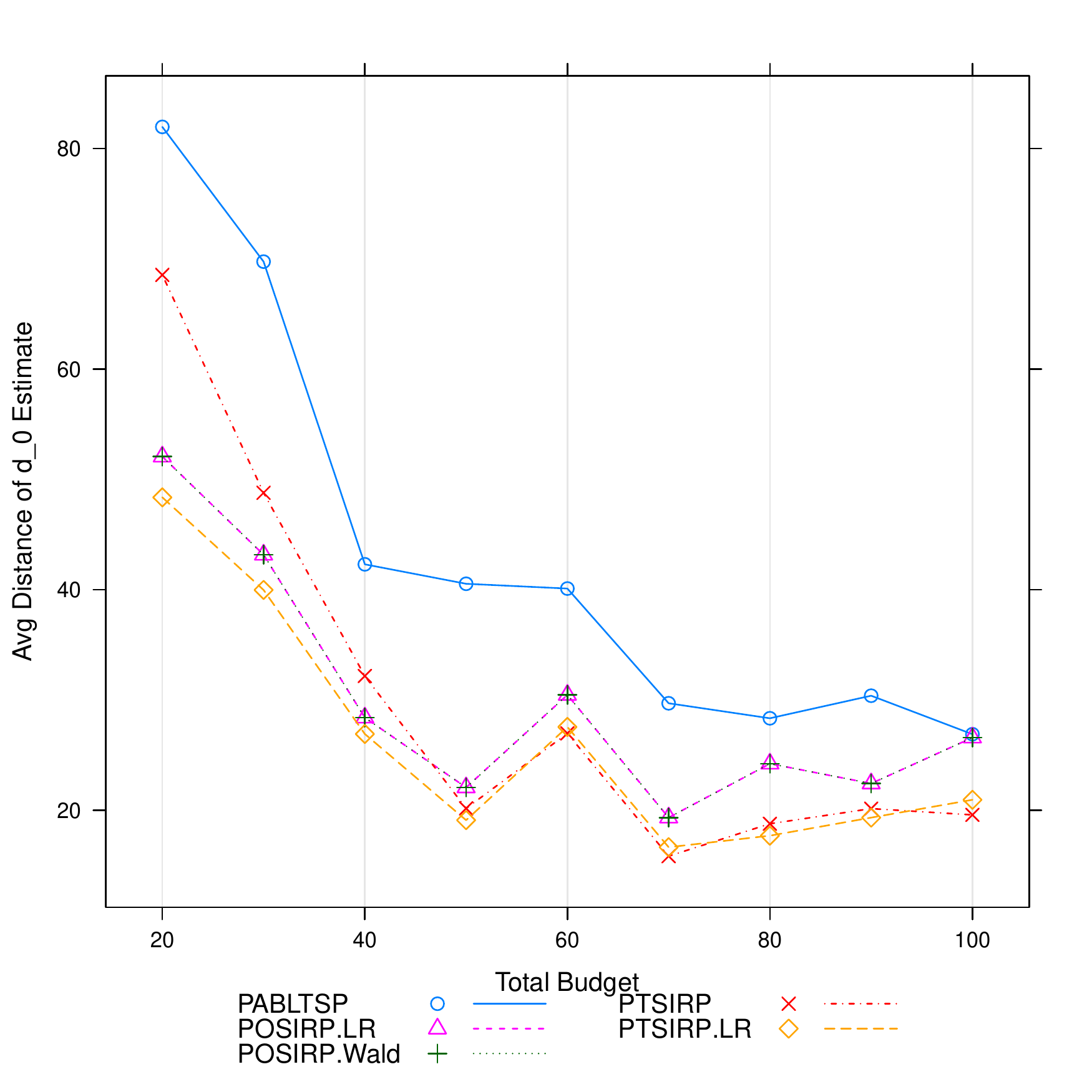}
\includegraphics[width=0.48\columnwidth]{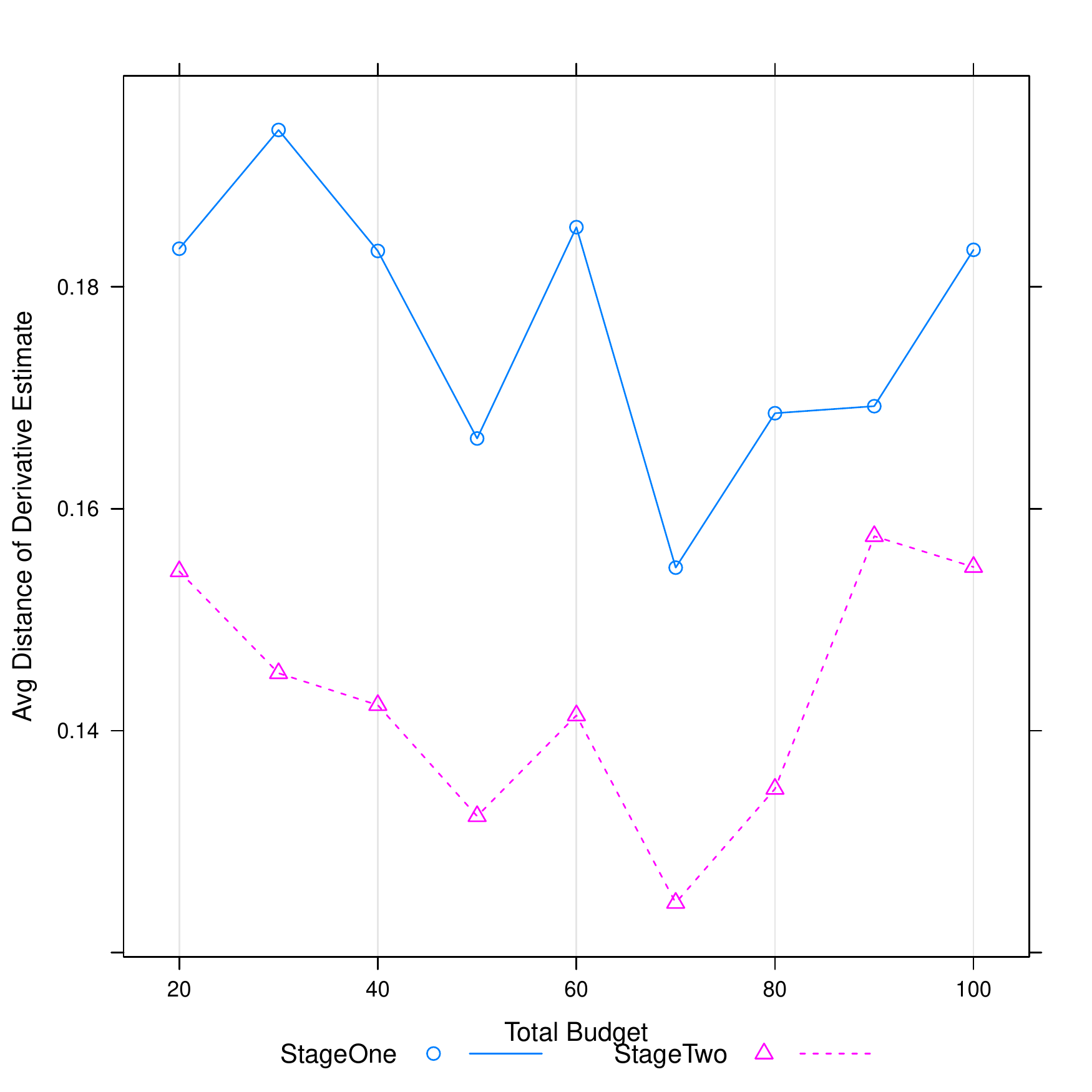}
\caption{Results obtained after combining both stage samples. The top panel shows the coverage rate and average length of confidence intervals generated by the five different procedures. The bottom panel shows the distance of the point estimate to the ``true'' value, and the distance of the derivative estimate to its ``true'' value.}
\label{fig:carExample:smalln:combined}
\end{center}
\end{figure}

Next, we perform another experiment to assess the reliability of the procedures with the FE data set.
We treat the data from the 1477 vehicles as the population, and sample from it according to different overall budgets of size $20,30,40,50,60,70,80,90,100$. 
Given the modest budgets, we combine samples from both stages for estimation of auxiliary parameters and inversion of the likelihood ratio. The results, averaged over 500 repetitions for each budget size, are depicted in Figure~\ref{fig:carExample:smalln:combined}.  Note that
POSIRP-Wald and PTSIRP-Wald still struggle to maintain coverage rates due to difficulties of auxiliary parameter estimation. 
The local linear approximation in PABLTSP also faces difficulties with such small budgets. 
On the other hand, POSIRP-LR and PTSIRP-LR maintain good coverage for all budgets. 
As noted before, PTSIRP-LR outperforms its one-stage counterpart with narrower intervals.
Considering the overall budgets investigated in this experiment, PTSIRP-LR performs well with an extremely small fraction of the overall data, illustrating its utility in the context of very large data sets, a topic of further discussion in the Discussion section.

\begin{figure}[!h]
  \begin{center}
  \includegraphics[width=.4\columnwidth,trim=0.4cm 0.85cm 0.4cm 1.45cm, clip=true]{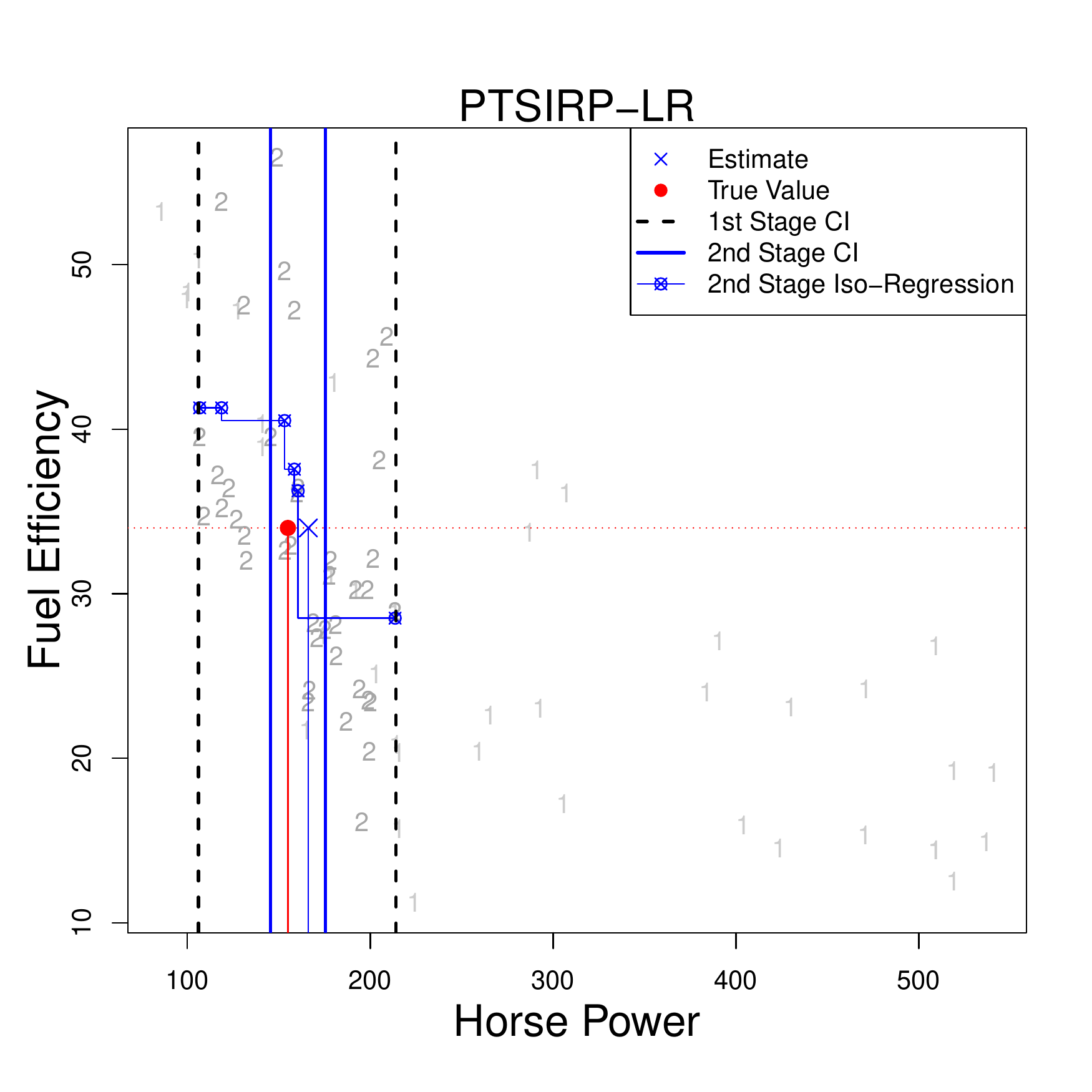}
  \caption
  [Plots for the data analysis 2016 CAFE results]
  {PTSIRP-LR results for estimating the 2016 CAFE standard of $d_0=m^{-1}(34)$.
   Numbers denote first and second stage samples, vertical lines denote corresponding confidence intervals, and `X' marks the final point estimate.}
  \label{paper:FG:DataAnalysis:PracticalProcedures:Cafe2016}
  \end{center}
\end{figure}

Finally, we return to the task discussed in the introductory section of estimating the horse power
at which the combined FE is equal to the upcoming 2016 CAFE standard of $d_0=m^{-1}(34)$.  
Employing isotonic regression on the entire sample yields a ``true" value of $d_0$ of around 155, 
with corresponding 95\% confidence interval [143.360,166.052]. Following the same procedure and budget allocations as above, PTSIRP-LR 
yields $\hat{d}_{0} = 166.204$ with corresponding 95\% confidence interval [145.599,175.450], as shown in Figure~\ref{paper:FG:DataAnalysis:PracticalProcedures:Cafe2016}. PTSIRP-LR covers the 
``true'' value with a reasonably sized interval, while utilizing a small fraction of the overall budget. 

The upshot of the analysis is that the two stage LR based procedure offers superior performance to
its competitors even with smaller budgets and when 
the underlying function exhibits nonlinearities. 

\section{Discussion and Concluding Remarks}
\label{TSPInvFun:paper:Discussion}
In this paper, we considered the estimation of 
the inverse of a monotone regression function
at a given point in a design setting. The results
strongly suggest that a two-stage procedure using
isotonic regression in both stages coupled with 
calculation of likelihood-ratio based confidence 
intervals is agnostic to the local structure in the vicinity
of the parameter of interest, requires minimal tuning and
exhibits superior performance.

The reader may wonder whether an alternative nonparametric procedure at stage one,
with a faster than the $n^{1/3}$ convergence rate of isotonic
regression may offer advantages to the proposed strategy.
We have investigated smoothed isotonic regression \citep{TangThesis2011} which, in a single stage, exhibits a convergence rate of $n^{2/5}$ and
when repeated in the second stage exhibits the same acceleration
pattern as isotonic regression provided the bandwidth is
appropriately chosen (for a detailed discussion of this 
subtle issue see \citep{TangThesis2011}). However, extensive numerical
work shows that no significant performance gains are realized, 
compared to using isotonic regression in both stages, while
at the same time a bandwidth parameter needs to be carefully
specified. Indeed, a strategy based on isotonic regression in the first
stage, followed by smooth isotonic regression in the second stage
struggles with the estimation of the iso-sine function presented in 
Figure 2.   

Finally, we should note that although the developed methodology
applies to design settings (where the investigator can select
the desired covariate and the corresponding response variable
values), it can also prove useful in the context of very large
data sets. Suppose that one is interested in estimating a
threshold of a monotone function from a very large data set
that can not be stored in its entirety in computer memory. In that case,
one-stage estimation based on the entire data set is computationally challenging, since it
requires appropriate modification of the standard algorithms. 
However, by adopting the proposed adaptive design framework,
one can overcome such computational difficulties, while still obtaining
a high degree of accuracy. It is our belief that by going to multiple stages, if necessary, with 
judiciously chosen parameters, one can match the performance of
the estimator based on all data that could be stored in
computer memory, thus providing a computationally efficient procedure that 
avoids major modifications of existing algorithms. The latter claim is supported by
the results of the experiment shown in Figure \ref{fig:carExample:smalln:combined}, 
and in Figure 4 of the Supplementary material, which 
indicates the PTSIRP-LR reduces computing time by substantial amounts at larger 
budgets compared to its one stage counterpart.


\makeatletter   
 \renewcommand{\@seccntformat}[1]{APPENDIX~{\csname the#1\endcsname}.\hspace*{1em}}
\makeatother

\appendix
\section{Proofs}
\label{TSPInvFun:paper:Appendix}
We discuss Propositions 1 and 2 of the paper. 

\subsection{Appendix of TSIRP}
\label{paper:AppendixForTSIRP}
We introduce the following \emph{idealized} two-stage isotonic regression procedure (ITSIRP) as follows:
\begin{enumerate}
  \item Set the first-stage sample proportion $p\in(0,1)$
  and let the first and second-stage sample sizes be
  $n_{1}=\lfloor np\rfloor$ and $n_{2}=n-n_{1}$, respectively,
  where $n$ is the total sample size.
  \item Let the ideal second-stage sampling interval be
  $[L_{1i}, U_{1i}]=[d_{0}\pm C_{1}n_{1}^{-\gamma_{1}}]$
  with $C_{1}>0$ and $\gamma_{1}>0$.
  \item Allocate the second-stage design points $\{X_{2,i}\}_{i=1}^{n_{2}}$
  according to a Lebesgue density $\tilde{g}_{2}$ on $[L_{1i}, U_{1i}]$, given by 
  $\tilde{g}_{2}(x)=(C_{1}n_{1}^{-\gamma_{1}})^{-1}\psi((x-d_{0})/(C_{1}n_{1}^{-\gamma_{1}}))$. 
  Denote the corresponding i.i.d. second-stage responses $\{Y_{2,i}\}_{i=1}^{n_{2}}$.
  \item Compute the unconstrained isotonic regression $m_{oI}$
        (and the constrained one $m_{oIc}$ under the null hypothesis $m^{-1}(\theta_{0})=d_{0}$) of $m$
        over $[L_{1i}, U_{1i}]$
        from the second-stage data.
  \item Obtain $d_{oI}=m_{oI}^{-1}(\theta_{0})$ and
        $2\log \lambda_{oI}= 2\log\lambda_{oI}(d_{0})
            = 2 \klm l_{n}(m_{oI}, \hat{\sigma})-l_{n}(m_{oIc}, \hat{\sigma}) \krm $,
        the ideal second-stage isotonic regression based estimator of $d_{0}$
        and likelihood ratio statistic 
        under $H_{0}: m^{-1}(\theta_{0})=d_{0}$; here, as before, $\hat{\sigma}$ is a consistent estimate of $\sigma$. 
\end{enumerate}
 {\bf Remark:} Note that ITSIRP is similar to TSIRP except that in ITSIRP the second-stage sampling interval is centered at $d_{0}$
instead of at $d_{1,I}$; therefore, the sampling density at Stage 2 in ITSIRP is $\psi$ renormalized to $[L_{1i}, U_{1i}]$, just as 
$g_{2}$ is $\psi$ renormalized to $[L_1, U_1]$ (see Proposition \ref{TSIRP:Prop:d2I:Wald}) in TSIRP. Since $d_{1,I}$ converges  to $d_{0}$ at rate $n^{1/3}$, which is 
faster than the rate at which $[L_1, U_1]$ is decreasing around $d_{1,I}$ (since $\gamma < 1/3$), 
$[L_1, U_1]$ is essentially indistinguishable from its idealized counterpart $[L_{1i}, U_{1i}]$ 
and the asymptotic behavior of $d_{oI}$ will be \emph{identical} to that of $d_{2I}$.  Similarly, the asymptotic 
distribution of the idealized LRT, $2\log \lambda_{oI}$, will be the \emph{same} as that of $2\log \lambda_{2,I}$. 

A rigorous proof of Proposition 1, formalizing the intuition above, can be provided via 
conditioning arguments similar in spirit to those used for proving Theorem 2 in \cite{Lan2007} that establishes the distributional convergence of the two-stage estimator of a change-point in a regression model; more specifically, the proof of Lemma 3.2 (a key intermediate step in proving Theorem 2) of a process convergence result proceeds by  conditioning on the values of the relevant estimates at Stage 1 in conjunction with some uniformity arguments. Proposition 2 requires similar conditioning strategies. In this paper, we provide a sketch of the derivations of the limiting distributions of the `idealized' (surrgoate) quantities $d_{oI}$ and $2\log\lambda_{oI}$. We first introduce the following quantities. 
\newline
\newline
For positive constants $a,b$ we define the process $X_{a,b}(t) \equiv a\,W(t) + b\,t^2$ where $W(t)$ is two-sided Brownian motion on $\mathbb{R}$, starting from 0. For a function $f$ defined on $\mathbb{R}$, let $\mbox{slogcm}\,(f,I)$ denote the left slope of the greatest convex minorant of the restriction of $f$ to the interval $I$. Define $g_{a,b}(t) = \mbox{slogcm}\,(X_{a,b}, \mathbb{R})(t)$ and $g_{a,b}^0(t) =  \{\mbox{slogcm}\,(X_{a,b}, (-\infty, 0])(t) \wedge 0\}\,1(t \leq 0) +  
\{\mbox{slogcm}\,(X_{a,b}, (0, \infty])(t) \vee 0\}\,1(t > 0)$. Define $\mathbb{D} = \int\,\{(g_{1,1}(t))^2 - (g_{1,1}^0(t))^2\}\,dt$ and recall that $\mathcal{Z} \equiv \argmin_{t}\,X_{1,1}(t)$ is the Chernoff random variable. 

\begin{TangThm}\label{TSIRP:Ideal:Thm:doI:Wald}
Under Assumption A, we have
  \begin{equation*}
n^{(1+\gamma_{1})/3}(d_{oI}-d_{0})
 \overset{d}{\rightarrow}
 C_{d_{oI}}
 \mathcal{Z},
  \end{equation*}
 where
 $C_{d_{oI}}= C_{d_{I}}\left(\frac{C_{1}}{(1-p)p^{\gamma_{1}}\psi(0)}\right)^{1/3}$.
\end{TangThm}

%
\begin{TangThm}\label{TSIRP:Ideal:Thm:d0:LRT}
  Under Assumption A and
  the null hypothesis $H_{0}: m^{-1}(\theta_{0})=d_{0}$,
$2\log\lambda_{oI} \overset{d}{\rightarrow} \mathbb{D}$.
\end{TangThm}


\begin{proof}[Proof-sketch of Theorem \ref{TSIRP:Ideal:Thm:doI:Wald}]
For every $x\in \mathbb{R}$,
\begin{eqnarray}
\label{lesstomore}
     P\left( n_{2}^{(1+\gamma_{1})/3}(d_{oI}-d_{0}) \leq x  \right)
       &=&  P\left( d_{oI} \leq d_{0} + xn_{2}^{-(1+\gamma_{1})/3} \right)  \nonumber \\
       &=&  P\left( \theta_{0} \leq m_{oI} (d_{0} + xn_{2}^{-(1+\gamma_{1})/3}) \right) \nonumber \\
       &=&  P\left( n_{2}^{(1+\gamma_{1})/3}( m_{oI}(d_{0} + xn_{2}^{-(1+\gamma_{1})/3})-\theta_{0}) \geq 0\right) \\
       &=&  P\left( n_{2}^{(1+\gamma_{1})/3}( m_{oI}(d_{0} + xn_{2}^{-(1+\gamma_{1})/3}) - m(d_{0}) ) \geq 0\right).
    \end{eqnarray}
 Thus, it is sufficient to derive the limiting distribution of
 $n_{2}^{(1+\gamma_{1})/3}( m_{oI}(d_{0} + xn_{2}^{-(1+\gamma_{1})/3}) - m(d_{0}) )$. 
  Deducing this limit involves three main steps: the first uses a switching relationship to change the original problem into
  an M-Estimation problem; the second solves the M-Estimation problem in the framework of empirical
  process theory; the third simplifies the final limit distribution. This approach is, by now, standard in dealing with the 
  asymptotics of isotonic estimates; see, for example, pages 296-299 of \cite{Vaart1996}. Without loss of generality, we 
  take $[a,b] = ]0,1]$ from here on. 
\newline
\newline
In the first step, we show
 \begin{TangLem}\label{Step1_Lemma_Swiching_Relationship}
For $t\in [d_{0}\pm C_{1}n_{1}^{-\gamma_{1}}]$ and $s\in \mathbb{R}$,
  \begin{equation}
    m_{oI}(t)\leq s
    \Leftrightarrow \underset{x\in [d_{0}\pm C_{1}n_{1}^{-\gamma_{1}}]}{argmin}
    \kll V_{n_{2}}(x)-sG_{n_{2}}(x)\krl
    \geq T(t),
  \end{equation}
where, 
\begin{equation}\label{Vn}
    V_{n_{2}}(x)=\frac{1}{n_{2}}\sum_{i=1}^{n_{2}}Y_{2,i}1(X_{2,i}\leq x),
    ~~
    G_{n_{2}}(x)=\frac{1}{n_{2}}\sum_{i=1}^{n_{2}}1(X_{2,i}\leq x);
\end{equation}
$T(t)$ is the largest $X_{2,i}$ less than or equal to $t$.
\end{TangLem}
This equivalence is called the `switching relationship' and can be derived by arguments similar to those leading to the last display on page 298 of van der Vaart and Wellner (1996). Hence, 
\begin{eqnarray*}
 & &P\left(n_{2}^{(1+\gamma_{1})/3}(m_{oI}(d_{0}+xn_{2}^{-(1+\gamma_{1})/3})-m(d_{0}))\leq z\right)\\
  &=& P\left(m_{oI}(d_{0}+xn_{2}^{-(1+\gamma_{1})/3}) \leq m(d_{0}) + zn_{2}^{-(1+\gamma_{1})/3}\right) \\
   &=& P\left(\underset{ x \in [d_{0}\pm C_1 n_{1}^{-\gamma_{1}}]}{argmin}
   \kll V_{n_{2}}(x)-(m(d_{0}) + zn_{2}^{-(1+\gamma_{1})/3})G_{n_{2}}(x)\krl
   \geq T(d_{0}+xn_{2}^{-(1+\gamma_{1})/3})\right).
\end{eqnarray*}
In the second step, by arguments similar to those on Page 299 of van der Vaart and Wellner (1996), we establish: 
\begin{TangLem}\label{Step2_theorem}
Under Assumption \textbf{(A)}, as $n\rightarrow \infty$,
\begin{eqnarray*}
   & & P\left(\underset{x \in [d_{0}\pm C_1 n_{1}^{-\gamma_{1}}]}{argmin}
   \kll V_{n_{2}}(x)-(m(d_{0}) + zn_{2}^{-(1+\gamma_{1})/3})G_{n_{2}}(x)\krl
   \geq T(d_{0}+xn_{2}^{-(1+\gamma_{1})/3})\right) \\
   &\rightarrow& P\left(\underset{h\in \mathbb{R}}{argmin}
   \kll X_{c,d} -zh \krl \geq x\right),
\end{eqnarray*}
where $c = \left(C_1\,\sigma^2\,((1-p)/p)^{\gamma}/\psi(0)\right)^{1/2}$ and  $d = m'(d_0)/2$. 
\end{TangLem}
In the third step, we use another switching relationship, namely: 
\begin{equation}\label{SwitchingRelationship2}
    g_{c,d}(x) > \lambda \Longleftrightarrow
    \underset{t\in\mathbb{R}}{argmin}(X_{c,d}(t)-\lambda t) < x,
    ~~
    \text{for} ~~ \lambda\in\mathbb{R},
\end{equation}
and the continuity of the random variables involved in the above display, to get: 
\begin{equation*}
  P\left( \underset{h\in \mathbb{R}}{argmin}
  \kll X_{c,d} - zh \krl \geq x\right)
  = P\left( g_{c,d}(x) \leq z \right).
\end{equation*}
Hence, 
\begin{equation} \label{loc-proc-conv}
n_2^{(1+\gamma_1)/3}\,( m_{oI}(d_{0} + xn_{2}^{-(1+\gamma_{1})/3}) - m(d_{0}) ) \rightarrow_d  g_{c,d}(x)
  \end{equation}
It follows from (\ref{lesstomore}) that 
\begin{equation}
\label{local-proc-conv-pr}
P\,( n_{2}^{(1+\gamma_{1})/3}(d_{oI} - d_0 ) \leq x) {\rightarrow} P(\,g_{c,d}(x) \geq 0) \,.
  \end{equation}
Then, using (\ref{SwitchingRelationship2}) again, we have
\begin{equation}\label{1}
    P\left( n_{2}^{(1+\gamma_{1})/3}(d_{oI}-d_{0}) \leq x  \right)
    \rightarrow P\left( \underset{t\in\mathbb{R}}{argmin}X_{c,d}(t) \leq x  \right).
\end{equation}
Now, from Problem 5 on Page 308 of van der Vaart and Wellner (1996), we have $\arg \min\,X_{c,d}(h) = (c/d)^{2/3}\,\arg \min X_{1,1}(t)$, whence 
\begin{equation}\label{3}
    P\left( n_{2}^{(1+\gamma_{1})/3}(d_{oI}-d_{0}) \leq x  \right)
    \rightarrow P\left( (c/d)^{2/3}\underset{t\in\mathbb{R}}{argmin}X_{1,1}(t) \leq x   \right).
\end{equation}
Since
$$(c/d)^{2/3}= \left(\frac{4\sigma^{2}}{m'(d_{0})^{2}}\cdot
\frac{(1-p)^{\gamma_{1}}C_{1}}{p^{\gamma_{1}}\psi(0)}\right)^{1/3},$$
 we have
 $$n_{2}^{(1+\gamma_{1})/3}(d_{oI}-d_{0})
 \overset{d}{\rightarrow} C_{d_{I}}\left(\frac{(1-p)^{\gamma_{1}}C_{1}}{p^{\gamma_{1}}\psi(0)}\right)^{1/3}
 \mathcal{Z},$$
 which leads to
$n^{(1+\gamma_{1})/3}(d_{oI}-d_{0})
 \overset{d}{\rightarrow} C_{d_{oI}}
 \mathcal{Z}$, the result in Theorem \ref{TSIRP:Ideal:Thm:doI:Wald}.
\end{proof}

\begin{proof}[Proof-sketch of Theorem \ref{TSIRP:Ideal:Thm:d0:LRT}]  
For simplicity, we assume the second-stage sampling density is uniform on $[L_{1i}, U_{1i}]$.
That is, $g_{2}(x)=(2C_{1}n_{1}^{-\gamma_{1}})^{-1}$ for $x\in[L_{1i}, U_{1i}]$.
Then, similar to (\ref{loc-proc-conv}), we have
\begin{equation}\label{ITSIRP:WeakConvergence:ConstrainedUnconstrained}
     \left(
       \begin{array}{c}
         n_{2}^{(1+\gamma_{1})/3}(m_{oI}(d_{0}+xn_{2}^{-(1+\gamma_{1})/3})-m(d_{0})) \\
         n_{2}^{(1+\gamma_{1})/3}(m_{oIc}(d_{0}+xn_{2}^{-(1+\gamma_{1})/3})-m(d_{0})) \\
       \end{array}
     \right)
     \overset{d}{\rightarrow}
     \left(
       \begin{array}{c}
         g_{c,d}(x) \\
         g_{c,d}^{o}(x) \\
       \end{array}
     \right),
\end{equation}
where now, $c=(2C_{1}\sigma^{2}[(1-p)/p]^{\gamma_{1}})^{1/2}$ and $d=m'(d_{0})/2$.
In fact,
the weak convergence
(\ref{ITSIRP:WeakConvergence:ConstrainedUnconstrained})
holds not only finite dimensionally,
but also in the normed linear space $L_{2}[-K,K] \times L_2[-K,K]$ for every $K > 0$,  because of the monotonicity of both $m_{oI}$ and $m_{oIc}$.

To derive the asymptotics for $2\log \lambda_{oI}$, it suffices (by Slustky's theorem) to consider a tweaked version of this quantity with the $\hat{\sigma}^2$ in the denominator replaced 
by the true $\sigma^2$. In what follows, we work with this version and continue to call it $2\log \lambda_{oI}$. We have: 
\begin{align*}
    2\log \lambda_{oI}
    & = 
    2\klm
    \frac{1}{2\sigma^{2}} \sum_{i=1}^{n_{2}} (Y_{2,i} - m_{oIc}(X_{2,i}))^{2}
    -
    \frac{1}{2\sigma^{2}} \sum_{i=1}^{n_{2}} (Y_{2,i} - m_{oI}(X_{2,i}))^{2}
    \krm \\
    & =
    \frac{1}{\sigma^{2}}\kll
    \sum_{i =1}^{n_2} \klm (Y_{2,i}-\theta_{0}) - (m_{oIc}(X_{2,i})-\theta_{0})\krm^{2}
    -
    \sum_{i =1}^{n_2} \klm (Y_{2,i}-\theta_{0}) - (m_{oI}(X_{2,i})-\theta_{0})\krm^{2}
    \krl  \\
    & =
    - \frac{2}{\sigma^{2}}
    \klm
    \sum_{i=1}^{n_{2}}(Y_{2,i}-\theta_{0})(m_{oIc}(X_{2,i})-\theta_{0})
    -
    \sum_{i=1}^{n_{2}}(Y_{2,i}-\theta_{0})(m_{oI}(X_{2,i})-\theta_{0})
    \krm \\
    & +
    \frac{1}{\sigma^{2}}\sum_{i=1}^{n_{2}}
    \klm
    (m_{oIc}(X_{2,i})-\theta_{0})^{2}
    -
    (m_{oI}(X_{2,i})-\theta_{0})^{2}
    \krm   \\
    & =
    - \frac{2}{\sigma^{2}}
    \sum_{i=1}^{n_{2}}(Y_{2,i}-m_{oIc}(X_{2,i}))(m_{oIc}(X_{2,i})-\theta_{0}) - \frac{2}{\sigma^2}\,\sum_{i=1}^{n_2}\,(m_{oIc}(X_{2,i}) - \theta_0)^2  \\
    & + \frac{2}{\sigma^{2}}
    \sum_{i=1}^{n_{2}}(Y_{2,i}-m_{oI}(X_{2,i}))(m_{oI}(X_{2,i})-\theta_{0}) +  \frac{2}{\sigma^2}\,\sum_{i=1}^{n_2}\,(m_{oI}(X_{2,i}) - \theta_0)^2 \\
    & +
    \frac{1}{\sigma^{2}}\sum_{i=1}^{n_{2}}
    \klm
    (m_{oIc}(X_{2,i})-\theta_{0})^{2}
    -  
    (m_{oI}(X_{2,i})-\theta_{0})^{2}
    \krm \\
    & =
    \frac{1}{\sigma^{2}}\sum_{i=1}^{n_{2}}
    \klm
    (m_{oI}(X_{2,i})-\theta_{0})^{2}
    -
    (m_{oIc}(X_{2,i})-\theta_{0})^{2}
    \krm,
\end{align*}
where the last equation is a consequence of the fact that isotonic regression estimators 
are formed by averaging the responses over blocks of order statistics of the covariates, which ensures that
\begin{equation*}
    \sum_{i=1}^{n_{2}}(Y_{2,i}-m_{oI}(X_{2,i}))(m_{oI}(X_{2,i})-\theta_{0})
    ~~~\text{and}~~~
    \sum_{i=1}^{n_{2}}(Y_{2,i}-m_{oIc}(X_{2,i}))(m_{oIc}(X_{2,i})-\theta_{0})
\end{equation*}
are both equal to 0.

Now denote $\mathbb{P}_{n_{2}}$ as the empirical measure of
the second-stage covariates $\{X_{2,i}\}_{i=1}^{n_{2}}$
and $P_{n_{2}}$ as the corresponding uniform probability measure
of $X_{2,i}$. Let $D_{n_{2}}$ denote the interval on which $m_{oI}$ and $m_{oIc}$ differ. 
Then, we have
\begin{equation*}
    2\log \lambda_{oI}
    =
    \frac{n_{2}}{\sigma^{2}}
    \mathbb{P}_{n_{2}}
    \klm
    (m_{oI}(x)-\theta_{0})^{2}
    -
    (m_{oIc}(x)-\theta_{0})^{2}
    \krm
    \{x\in D_{n_{2}}\}
    = T_{1} + T_{2},
\end{equation*}
where
\begin{align*}
    T_{1} & =
    \frac{n_{2}}{\sigma^{2}}
    (\mathbb{P}_{n_{2}} - P_{n_{2}})
    \klm
    (m_{oI}(x)-\theta_{0})^{2}
    -
    (m_{oIc}(x)-\theta_{0})^{2}
    \krm
    \{x\in D_{n_{2}}\},\\
    T_{2} & =
    \frac{n_{2}}{\sigma^{2}}
    P_{n_{2}}
    \klm
    (m_{oI}(x)-\theta_{0})^{2}
    -
    (m_{oIc}(x)-\theta_{0})^{2}
    \krm
    \{x\in D_{n_{2}}\}.
\end{align*}

Since $({1-2\gamma_{1})/3}<1/2$,
both $(m_{oI}(x)-\theta_{0})$
and $(m_{oIc}(x)-\theta_{0})$
are $O_{P}(n_{2}^{\frac{1+\gamma_{1}}{3}})$ and
\begin{equation*}
    T_{1} =
    \frac{1}{\sigma^{2}}
    n_{2}^{\frac{1-2\gamma_{1}}{3}}
    (\mathbb{P}_{n_{2}} - P_{n_{2}})
    \kll
    \klm n_{2}^{\frac{1+\gamma_{1}}{3}}(m_{oI}(x)-\theta_{0})\krm^{2}
    -
    \klm n_{2}^{\frac{1+\gamma_{1}}{3}}(m_{oIc}(x)-\theta_{0})\krm^{2}
    \krl
    \{x\in D_{n_{2}}\},
\end{equation*}
we can show that $T_{1}$ converges to 0 in probability
by empirical process theory arguments.

Next, $T_{2}$ is given by 
\begin{align*}
    & \frac{n_{2}}{\sigma^{2}}
    \int_{D_{n_{2}}}
    \klm
    (m_{oI}(x)-\theta_{0})^{2}
    -
    (m_{oIc}(x)-\theta_{0})^{2}
    \krm
    \frac{n_{1}^{\gamma_{1}}}{2C_{1}}dx \\
    = &
    \frac{1}{2C_{1}\sigma^{2}}
    \kls \frac{p}{1-p}\krs^{\gamma_{1}}
    n_{2}^{1+\gamma_{1}}
    \int_{D_{n_{2}}}
    \klm
    (m_{oI}(x)-\theta_{0})^{2}
    -
    (m_{oIc}(x)-\theta_{0})^{2}
    \krm dx \\
    = &
    \frac{1}{c^{2}}
    n_{2}^{1+\gamma_{1}}
    \int_{D_{n_{2}}}
    \klm
    (m_{oI}(x)-\theta_{0})^{2}
    -
    (m_{oIc}(x)-\theta_{0})^{2}
    \krm dx \\
    = &
    \frac{1}{c^{2}}
    \int_{n_{2}^{\frac{1+\gamma_{1}}{3}}(D_{n_2} - d_0)}
    \kll
    \klm
    n_{2}^{\frac{1+\gamma_{1}}{3}}(m_{oI}(d_{0}+tn_{2}^{-\frac{1+\gamma_{1}}{3}})-\theta_{0})
    \krm^{2}
    -
    \klm
    n_{2}^{\frac{1+\gamma_{1}}{3}}(m_{oIc}(d_{0}+tn_{2}^{-\frac{1+\gamma_{1}}{3}})-\theta_{0})
    \krm^{2}
    \krl dt \\
    \overset{d}{\rightarrow} &
    \frac{1}{c^{2}}
    \int
    \klm
    g_{c,d}(t)^{2}
    -
    g_{c,d}^{o}(t)^{2}
    \krm
    dt = \mathbb{D}.
\end{align*}
The equality preceding the weak convergence above follows from
the change of variable $x = d_{0} + tn_{2}^{-(1+\gamma_{1})/3}$
and the weak convergence of the likelihood ratio statistic follows from the weak convergence result
(\ref{ITSIRP:WeakConvergence:ConstrainedUnconstrained})
in the $L_{2}$ sense and the fact that the set $n_2^{\frac{1+\gamma_{1}}{3}}(D_{n_2} - d_0)$ is contained in a compact set with (arbitrarily) high 
probability, eventually.  For the very last equality, see, for example, the proof of Theorem 2.2 of \cite{Banerjee2007d}. 
Thus, Theorem \ref{TSIRP:Ideal:Thm:d0:LRT} holds.
 
\end{proof}

\bibliographystyle{ECA_jasa}
\bibliography{references}

\end{document}


\title {Two-Stage Plans for Estimating a Threshold Value of a Regression Function: Supplemental Technical Material}
\author[1]{Runlong Tang}
\author[2]{Moulinath Banerjee}
\author[2]{George Michailidis}
\author[2]{Shawn Mankad}
\affil[1]{Department of Operations Research and Financial Engineering, Princeton University}
\affil[2]{Department of Statistics, University of Michigan}
\renewcommand\Authfont{\scshape\small}
\renewcommand\Affilfont{\itshape\small}
\setlength{\affilsep}{0em}
\maketitle

\label{TSPInvFun:paper:SuppProofs}

\section{Multistage adaptive procedure and its convergence rate}
\label{paper:AppendixIRDisc}

Next, we discuss whether the parametric convergence rate of $\sqrt{n}$ can be achieved by using a multistage sampling procedure (with more than 2 stages) that constructs an isotonic regression estimate of $d_{0}$ at each stage. So, consider the generic adaptive procedure described at the beginning of 
Section 2.2 where we obtain $d_{2,I}$ by IR at Step 4 but instead of finding a confidence interval in Step 5, we select a neighborhood of $d_{2,I}$, 
say $[L_2, U_2]$, and continue sampling at Stage 3. Of course, in this case, we allocate our budget in proportions $p_1, p_2, p_3$ adding up to 1. Now, the convergence rate of $d_{2,I}$ for $d_0$ is $n^{(1+\gamma_1)/3}$ and $[L_2, U_2]$ is therefore chosen to be of the form $[d_{2,I} \pm C_2 n_2^{-\gamma_2}]$, 
with $\gamma_2 < (1+\gamma_1)/3$ and $n_2 \equiv n\,p_2$. Since $\gamma_1 < 1/2$, we have $\gamma_2 < 1/2$. Finally, $n_3 \equiv n\,p_3$ covariate-response 
pairs are sampled from $[L_2, U_2]$ at Stage 3 and the IR procedure is used to come up with a final estimate $d_{3,I}$ with convergence rate $n^{(1+\gamma_2)/3}$. But, as $(1+\gamma_2)/3 < 1/2$, this is again slower than $\sqrt{n}$. Following this line of argument, it is not difficult to see that no $k$-stage procedure based on 
IR at each stage can produce an estimator of $d_{0}$ that achieves the parametric rate. Note, also, that we can come as close as possible to $\sqrt{n}$ if $k$ is chosen 
large enough. A $k$-stage procedure involves a sequence $(\gamma_1, \gamma_2, \ldots > \gamma_{k-1})$ with $1/2 > \gamma_1$ and $\gamma_{i+1} < 
(1+\gamma_{i})/3$, for $i \geq 1$ and yields a final rate of convergence given by $(1+\gamma_{k-1})/3$. Now, take some large $k$  and consider a procedure where $(1+\gamma_{i-1})/3 > \gamma_i > (1+\gamma_{i-1})/3 - \eta$ for some (very small) $\eta > 0$, for $i = k, (k-1), \ldots, 2$. Then, using the second inequality time and again, by simple algebra:
\[ \frac{(1+\gamma_{k-1})}{3} > (1-\eta)\sum_{j=1}^{k-2}\,\left(\frac{1}{3}\right)^j + \left(\frac{1}{3}\right)^{k-1} + \frac{\gamma_{1}}{3^{k-1}} \,,\]
which can clearly be made as close to $1/2$ as one pleases for small (enough) $\eta$ and large (enough) $k$.

\section{Supplemental Material for Performance Evaluation of the Adaptive Procedures}
\label{paper:SuppForSection3}
In this section we will present additional results regarding the estimation of $m'(d_{0})$.

\begin{figure}
  \begin{center}
  \includegraphics[scale=0.65]{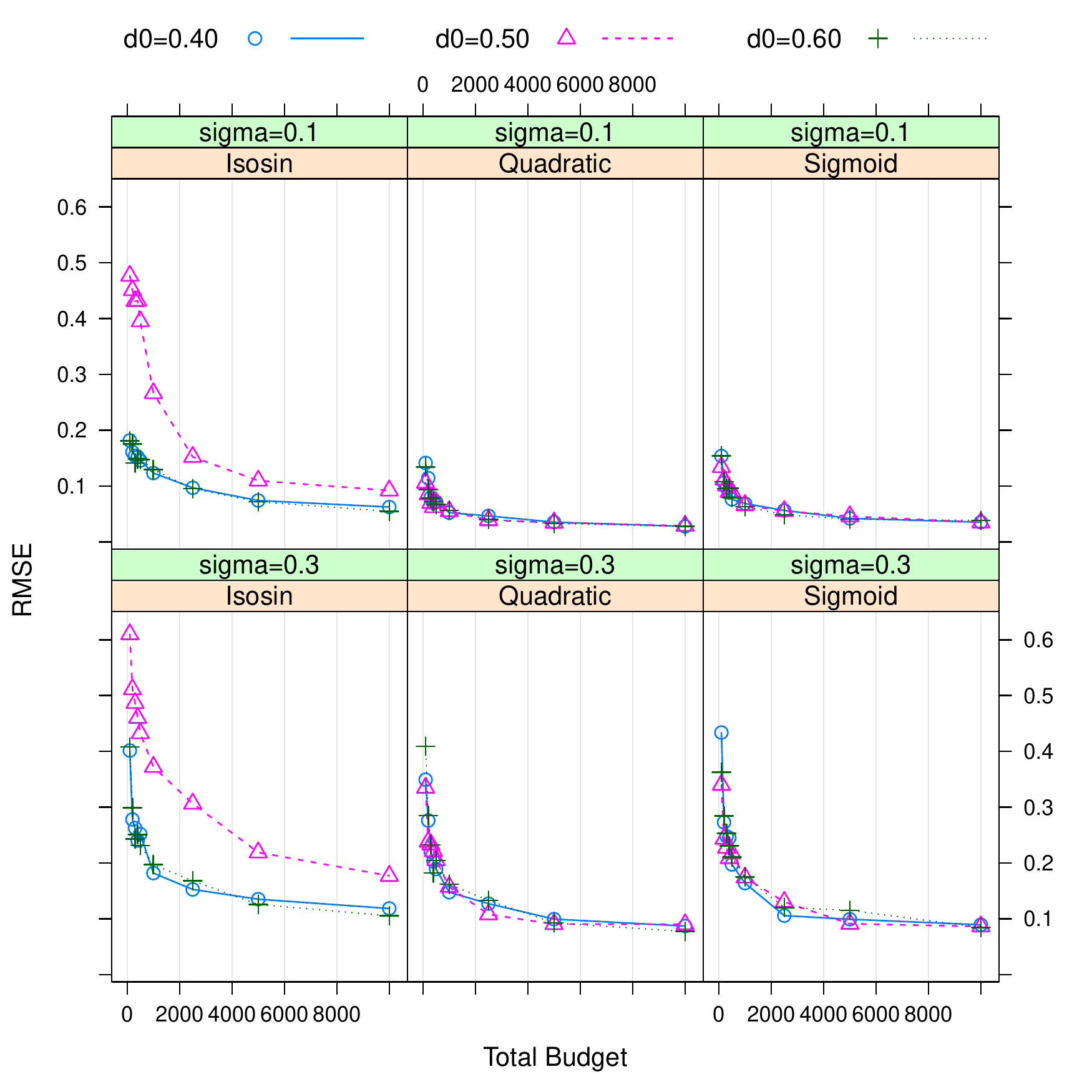}
  \caption
  [Root mean squared errors of $m'(d_0)$]
  {This figure shows the root mean squared errors of the estimates of $m'(d_0)$ using the local quadratic regression procedure proposed in \cite{Fan1996}. The first five data points correspond to sample sizes considered, e.g., sample sizes of 100,200,...,500.}
  \label{paper:FG:MSE:deriv}
  \end{center}
\end{figure}

We use a local quadratic regression procedure 
to estimate $m'(d_{0})$. 
An asymptotically optimal bandwidth that minimizes 
the asymptotic MSE is employed for this purpose. As expected and shown in Figure \ref{paper:FG:MSE:deriv}, 
the estimator tends to perform well with very large sample sizes. However, for the sample sizes considered in our numerical work, 
the performance is unsatisfactory especially 
under the isotonic sine function. The root mean squared error can be substantial, which causes 
the coverage rates reported in the main text to behave erratically.

\begin{figure}
  \begin{center}
  \includegraphics[scale=0.45]{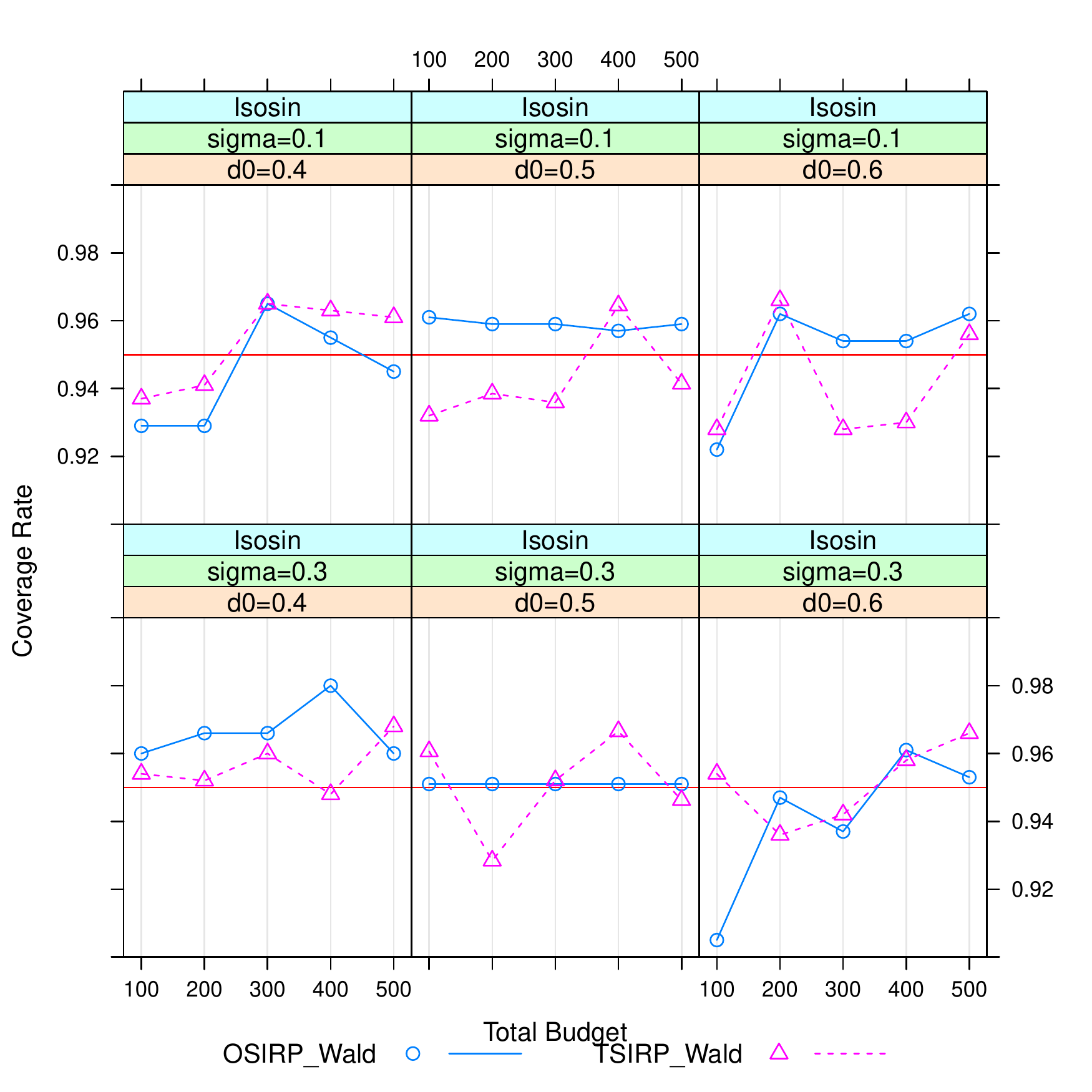}
  \includegraphics[scale=0.45]{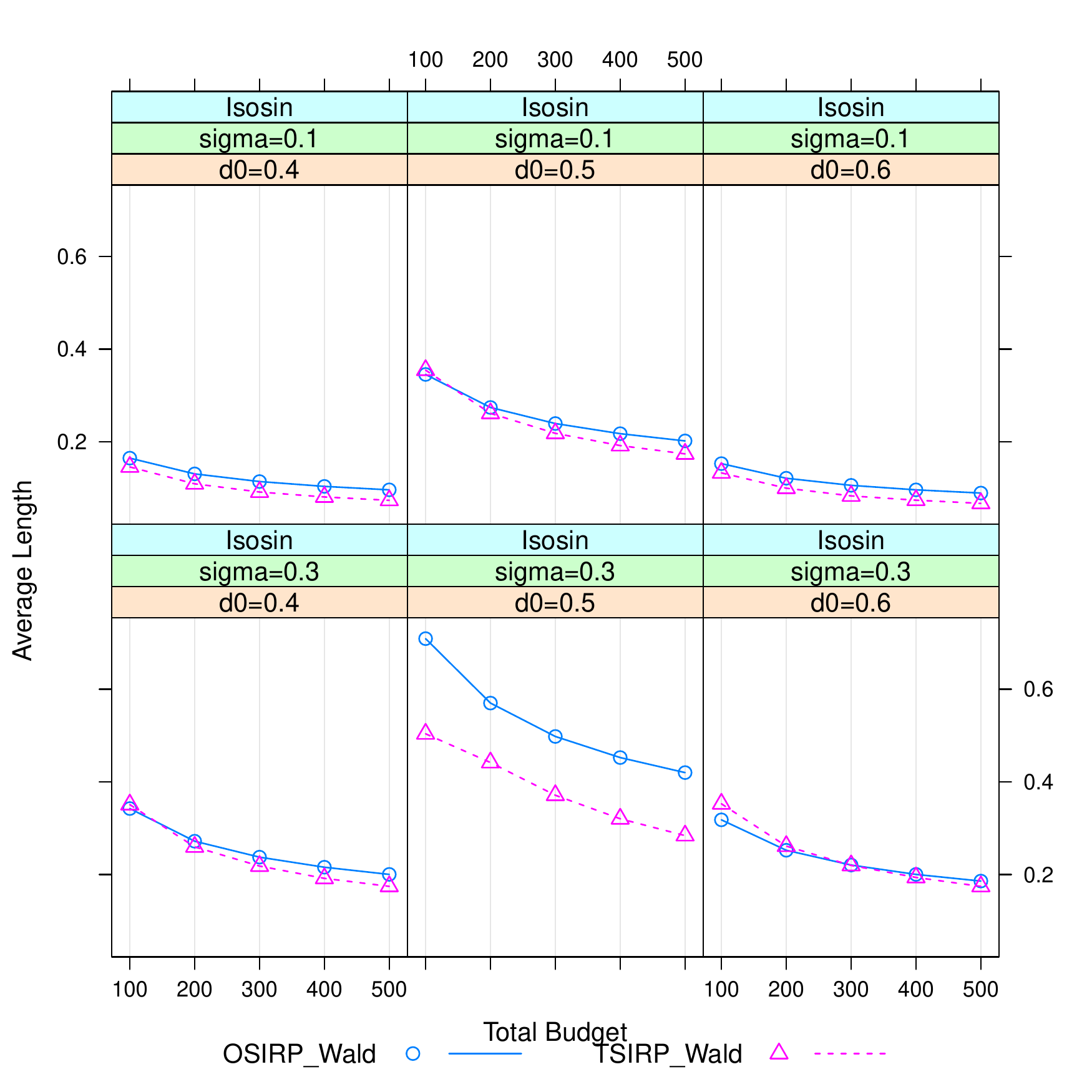}\\
  \caption
  [Coverage rates and average lengths with the isotonic sine function]
  {The left panel shows the coverage rates
of the 95\% confidence intervals for $d_{0}=0.5$ from the 
one and two stage IR-wald procedures
with the isotonic sine functions
and different values of $\sigma$ and $n$.
The right panel shows the average lengths
of the 95\% confidence intervals for $d_{0}$.
The derivative $m'(d_0)$ and $\sigma$ are assumed known.
}
  \label{paper:FG:isotonic sine:CRAndAL_true}
  \end{center}
\end{figure}

If we repeat the procedures with perfect knowledge of the nuisance parameter
$m'(d_0)$, then coverage rates are about the nominal level of $95\%$ (Figure \ref{paper:FG:isotonic sine:CRAndAL_true}). Therefore,
when the underlying regression function is well-behaved,
we can use the more aggressive PTSIRP. Otherwise,
we use the conservative but stable PTSIRP-LR, which avoids derivative estimation.

\section{Different Budget Allocations for Fuel Efficiency Analysis}

\begin{figure}[!h]
  \begin{center}
  \begin{tabular}{lccc}
  \raisebox{13ex}{\small $n=80$} &
  \includegraphics[width=.29\columnwidth, trim=0.4cm 0.85cm 0.4cm 1.45cm, clip=true]{OneStageIRWald_budget80.pdf} &
  \includegraphics[width=.29\columnwidth, trim=0.4cm 0.85cm 0.4cm 1.45cm, clip=true]{OneStageLR_budget80.pdf}&\\
  \raisebox{13ex}{\begin{tabular}{l}\small $n_{1}=35$\\\small $n_{2}=45$\end{tabular}} &
  \includegraphics[width=.29\columnwidth, trim=0.4cm 0.85cm 0.4cm 1.45cm, clip=true]{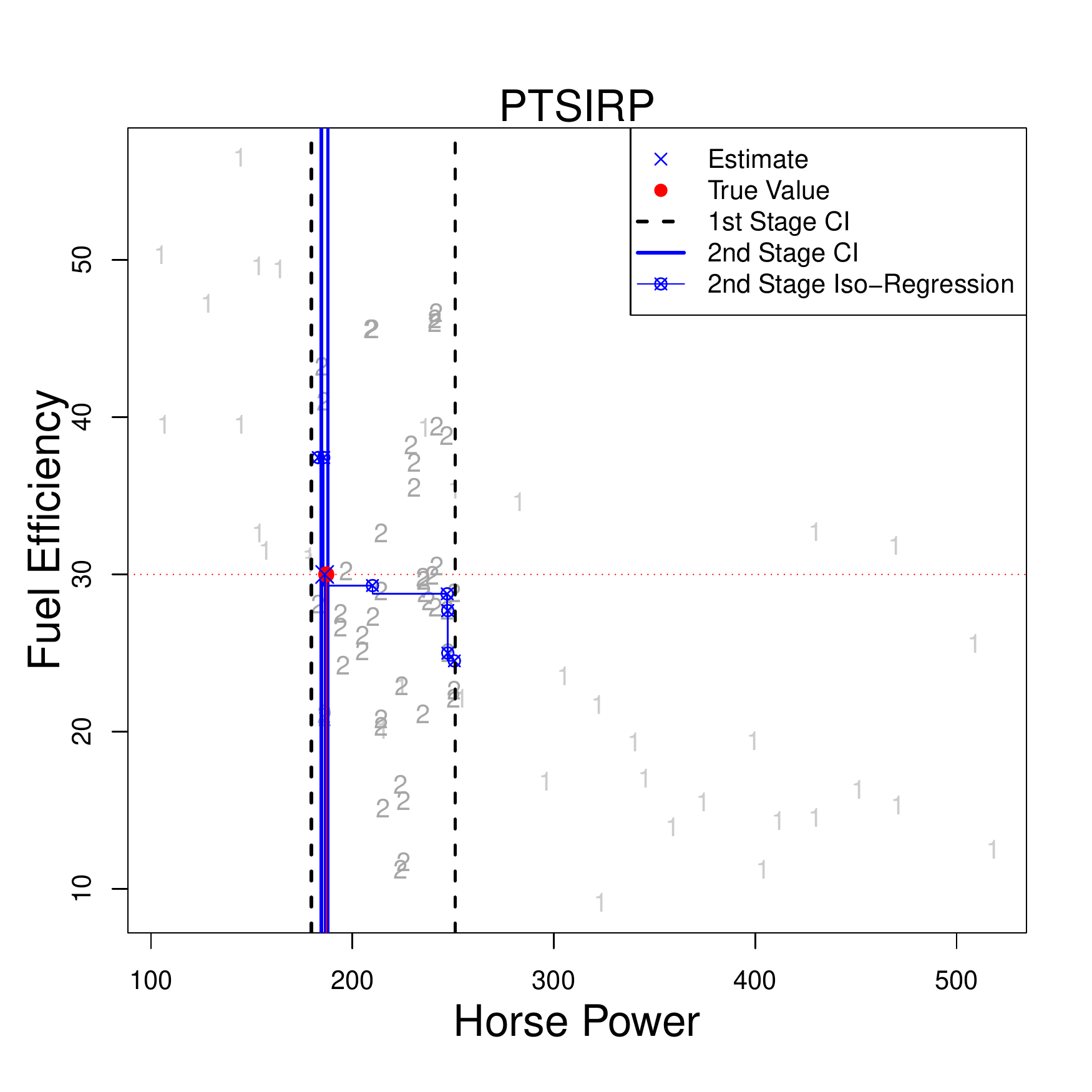} & 
    \includegraphics[width=.29\columnwidth, trim=0.4cm 0.85cm 0.4cm 1.45cm, clip=true]{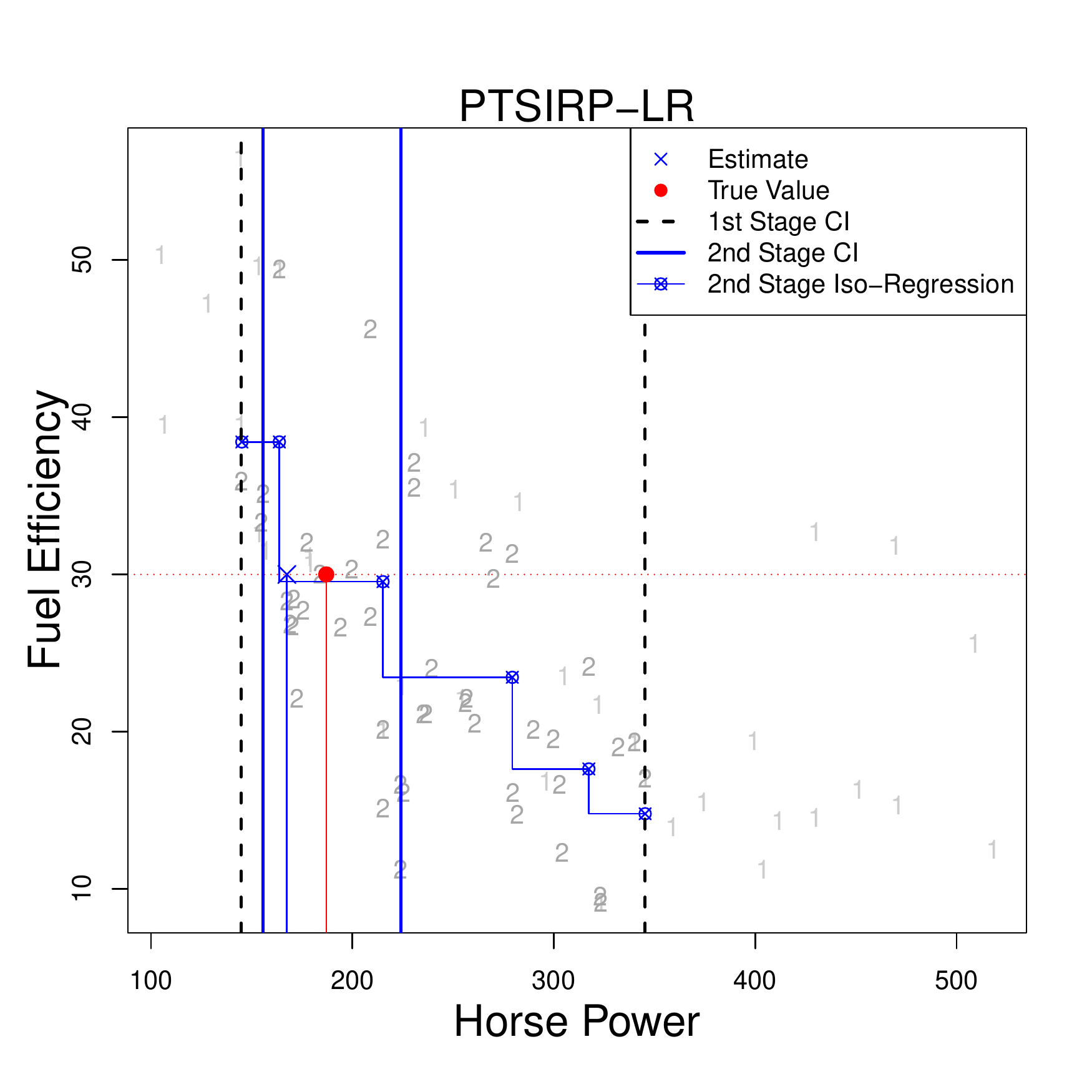} & 
  \includegraphics[width=.29\columnwidth, trim=0.4cm 0.85cm 0.4cm 1.45cm, clip=true]{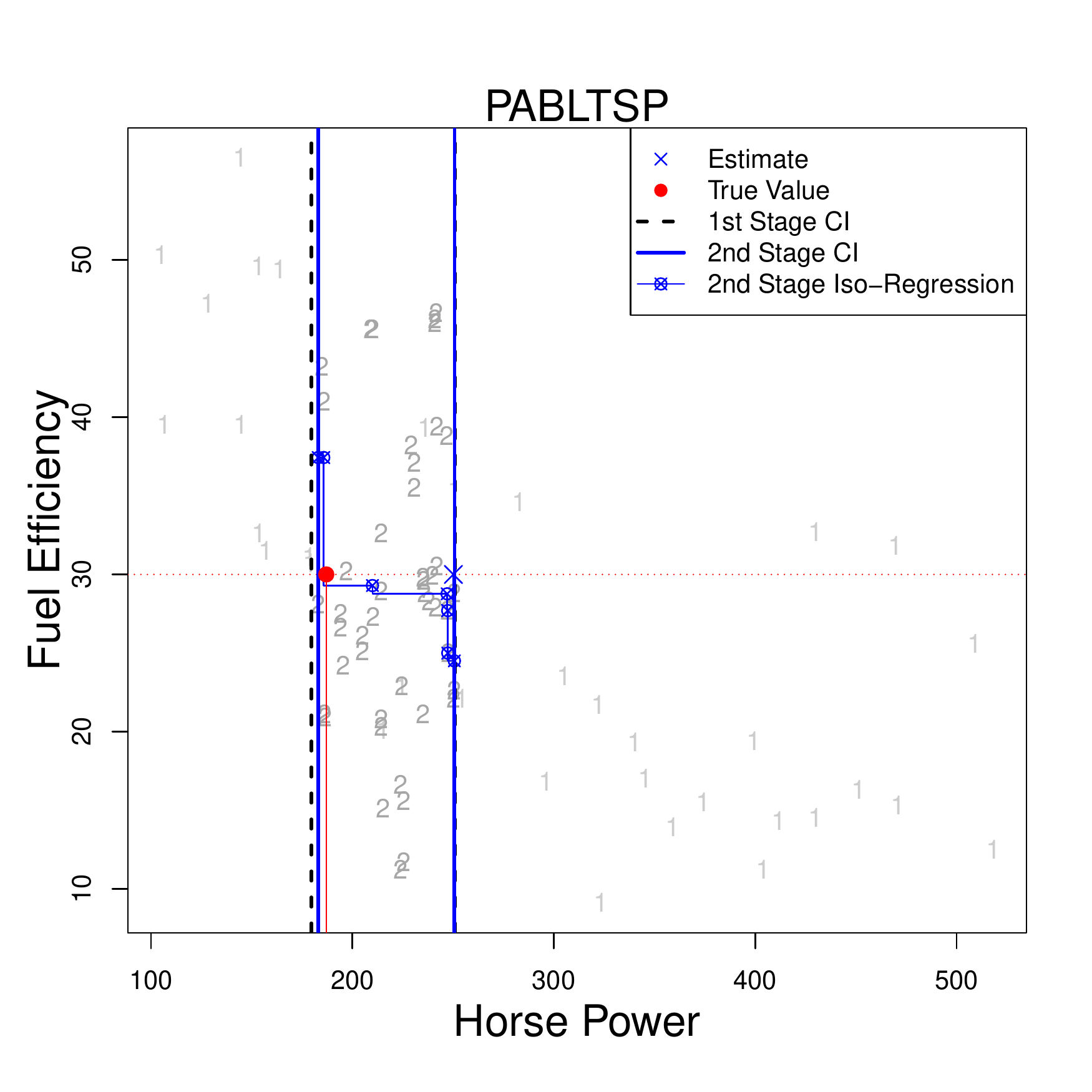} \\
  \raisebox{13ex}{\begin{tabular}{l}\small $n_{1}=40$\\\small $n_{2}=40$\end{tabular}} &
  \includegraphics[width=.29\columnwidth, trim=0.4cm 0.85cm 0.4cm 1.45cm, clip=true]{TwoStageIRIRWald_budget80_n1_40_n2_40.pdf} &
    \includegraphics[width=.29\columnwidth, trim=0.4cm 0.85cm 0.4cm 1.45cm, clip=true]{TwoStageLR_budget80_n1_40_n2_40.pdf} &  
  \includegraphics[width=.29\columnwidth, trim=0.4cm 0.85cm 0.4cm 1.45cm, clip=true]{TwoStageIRIRLocLinear_budget80_n1_40_n2_40.pdf} \\
  \raisebox{13ex}{\begin{tabular}{l}\small $n_{1}=45$\\\small $n_{2}=35$\end{tabular}} &
  \includegraphics[width=.29\columnwidth, trim=0.4cm 0.85cm 0.4cm 1.45cm, clip=true]{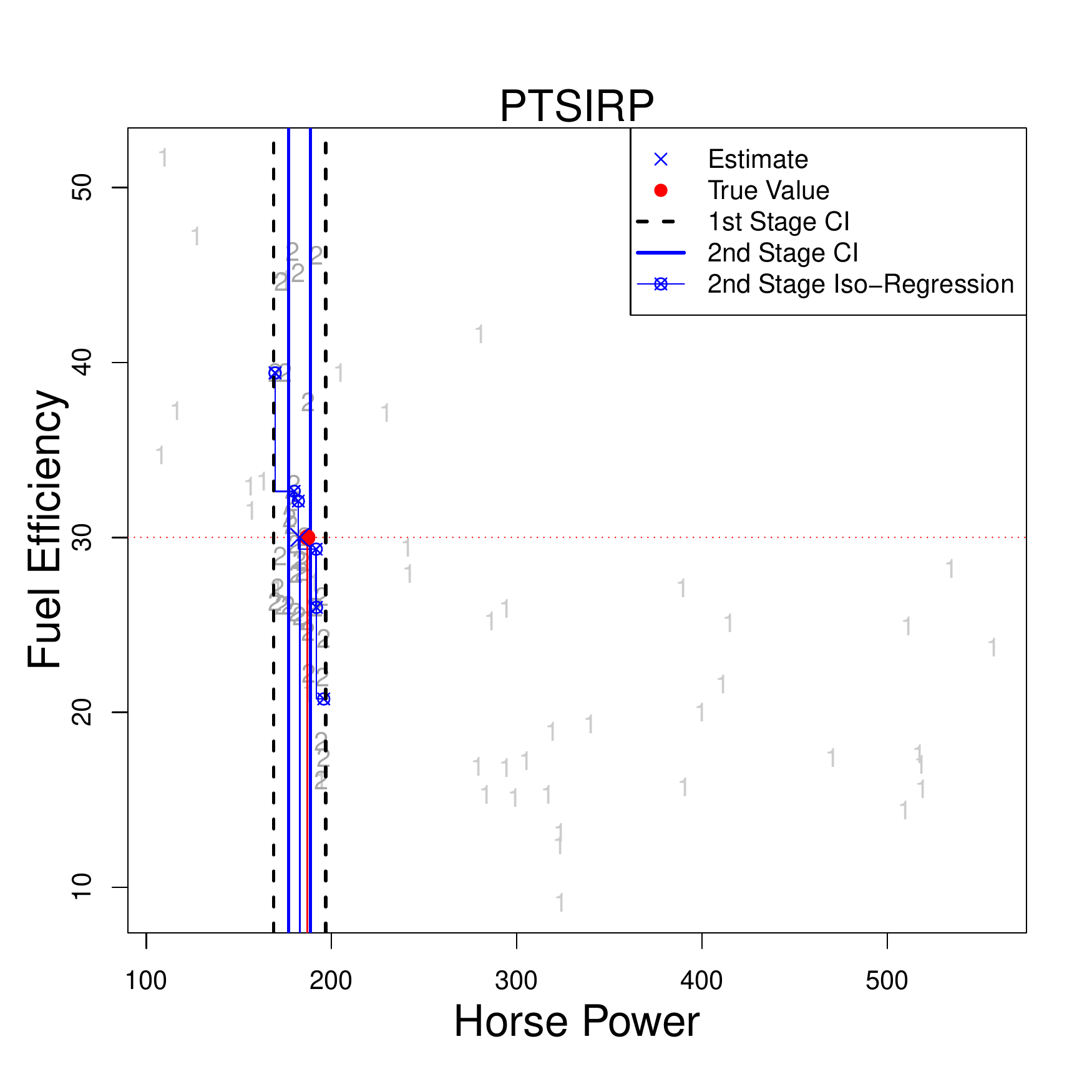} & 
    \includegraphics[width=.29\columnwidth, trim=0.4cm 0.85cm 0.4cm 1.45cm, clip=true]{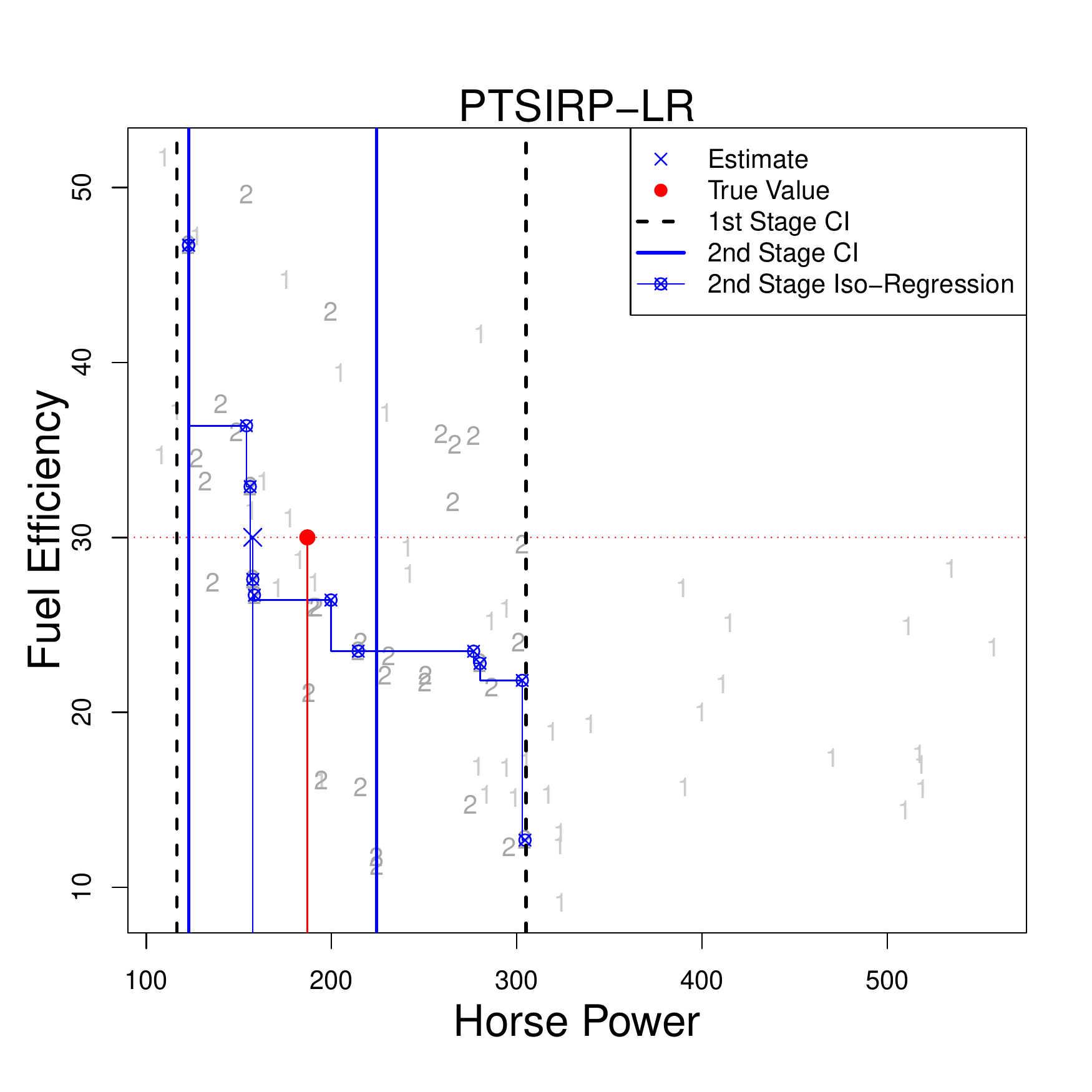} & 
  \includegraphics[width=.29\columnwidth, trim=0.4cm 0.85cm 0.4cm 1.45cm, clip=true]{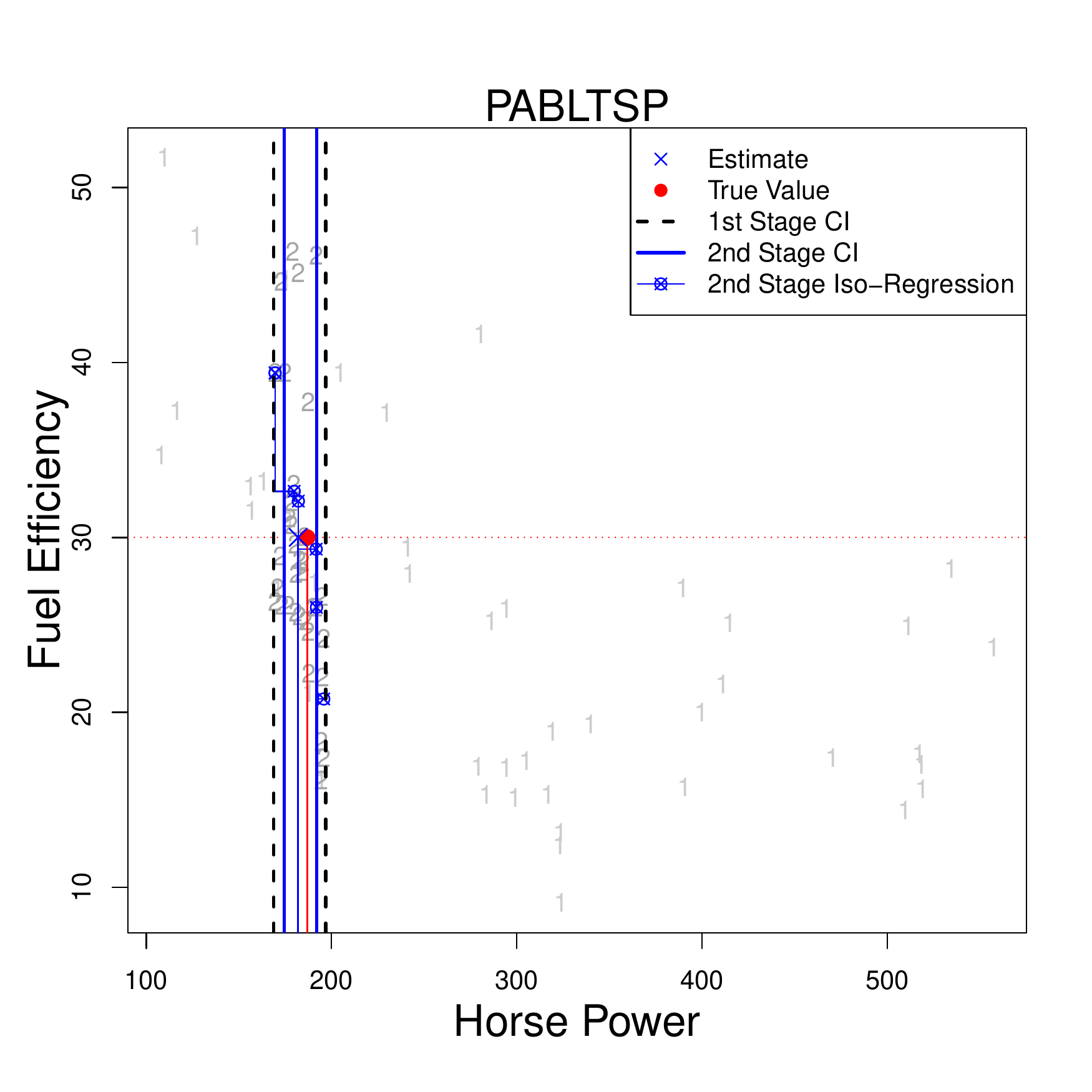} 
  \end{tabular}
  \caption
  [Plots for the data analysis results]
  {Plots for the data analysis with different budget specifications.}
  \label{paper:FG:DataAnalysis:PracticalProcedures:Budget80}
  \end{center}
\end{figure}

In this section, we repeat the analysis from the main text with different budget specifications.
In particular, we present in Figure \ref{paper:FG:DataAnalysis:PracticalProcedures:Budget80} 
three different scenarios maintaining the total budget of $80$ samples. 

Note that the one-stage procedures tend to provide poor point estimates. 
Whereas, the LR based two
stage procedure PTSIRP-LR and the local linear approximation (PABLTSP) tend to cover the 
``true'' value of $187$ in almost all cases with better point estimates. 

\begin{figure}
\centerline{
\includegraphics[width=.7\columnwidth]{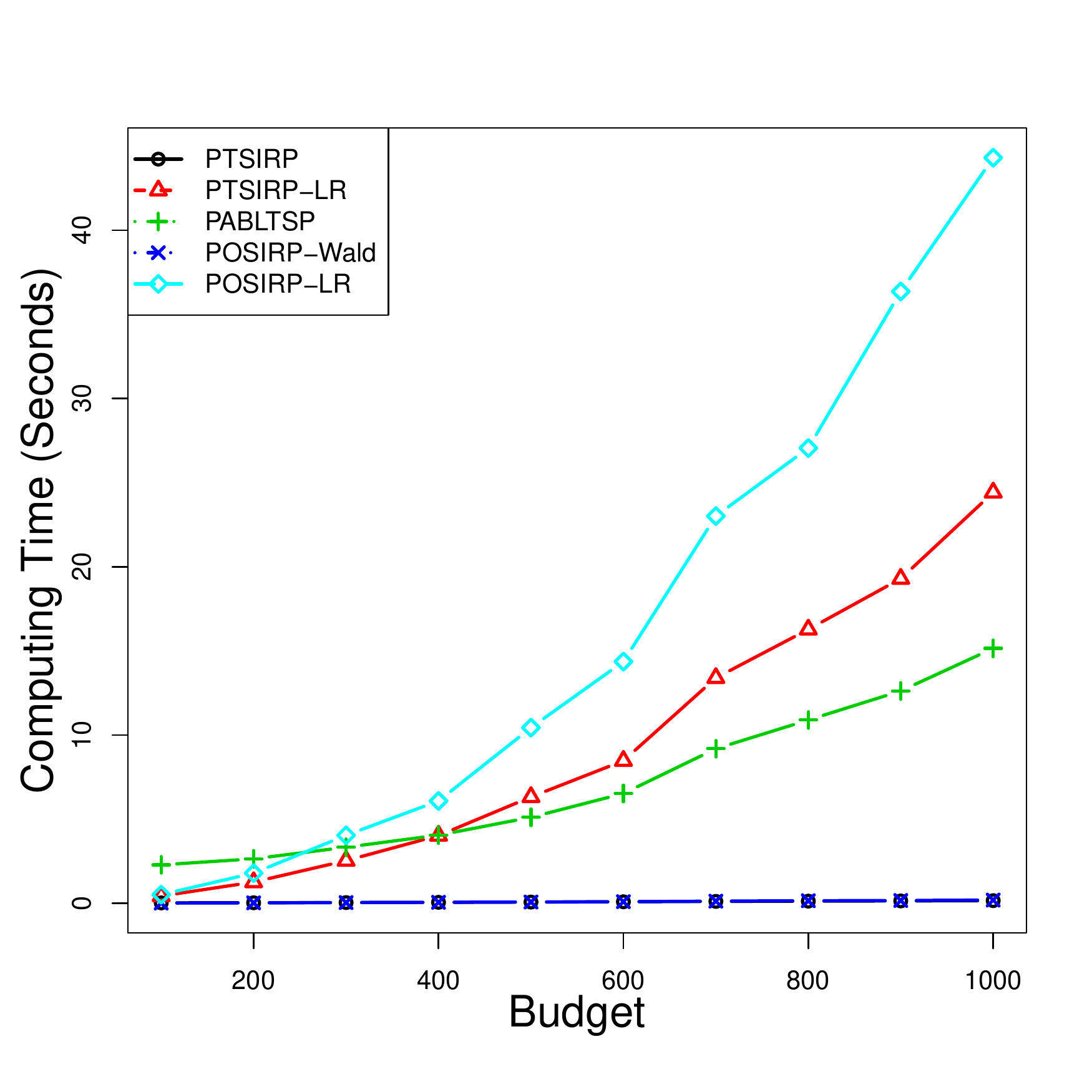} 
}
\caption{Average computing times with different budgets for the isotonic sine example.}
\label{fig:isosine:times}
\end{figure}

In addition to providing reliable estimates with smaller budgets and `ill-behaved' functions, PTSIRP-LR reduces computing time by substantial amounts at larger budgets compared to its one stage counterpart. Figure~\ref{fig:isosine:times} shows computing times, averaged over 500 trials, with 50\% of the overall budget allocated to the first stage for every trial. The Wald and local linear procedures can be performed faster, though their efficacy is dubious with smaller budgets and with `ill-behaved' functions due to auxiliary parameters estimation.

Altogether, our numerical studies have shown that likelihood ratio based procedures are robust, but do require inversion of the likelihood ratio, which adds computing cost. Utilizing two stages reduces computing time over POSIRP-LR, while also obtaining tighter confidence regions.

\bibliographystyle{ECA_jasa}
\bibliography{references}

%
%
%
%
%
